\documentclass[10pt,twocolumn,superscriptaddress,showpacs,prd,aps,amsmath,amssymb,nofootinbib]{revtex4-1}
\usepackage{graphicx,color,xcolor}
\usepackage{soul}
\usepackage{amsmath,amssymb}
\usepackage{verbatim}
\usepackage{float}
\usepackage{wasysym}
\usepackage{epsfig}
\usepackage{psfrag}
\usepackage{dsfont}
\usepackage{amsfonts}
\usepackage{mathrsfs}
\usepackage{multirow}
\usepackage{times}
\usepackage{bm}
\usepackage{scalefnt}
\usepackage{xspace}
\usepackage{mathtools}
\usepackage{lipsum}

\usepackage{blindtext, rotating}
\usepackage{pifont}
\usepackage[normalem]{ulem}   
\usepackage{grffile}
\usepackage{booktabs}
\usepackage{academicons}
\usepackage{orcidlink}
\usepackage{hyperref}
\usepackage{nameref}
\hypersetup{
  colorlinks=true,        
  linkcolor=blue,         
  citecolor=cyan,         
}
\graphicspath{{./}}

\newcommand{\Msol}{M_\odot}

\newcommand{\AT}{A_{T}}
\newcommand{\AP}{A_{P}}
\newcommand{\rone}{r_1}
\newcommand{\expfac}{e^{-(r/\rone)^{2p}}}

\newcommand{\cmark}{\ding{51}}%
\newcommand{\xmark}{\ding{55}}%

\begin{document}
\title{Impact of Magnetic Field Topology on Electromagnetic and Gravitational Waves from Binary Neutron Star Merger Remnants}
\author{Inês Rainho\orcidlink{0000-0003-4937-0638}}
\email{ianrain@uv.es}
\thanks{she/her}
\affiliation{Departamento de Astronom\'{\i}a y Astrof\'{\i}sica, Universitat de Val\`encia, Avinguda Vicent Andrés Estellés 19, 46100, Burjassot (Val\`encia), Spain}
\author{Jamie Bamber\orcidlink{0000-0001-7181-3365}}
\affiliation{Department of Physics, University of Illinois at Urbana-Champaign, Urbana, IL 61801, USA} 
\author{Davide Guerra\orcidlink{0000-0003-0029-5390}}
\affiliation{Departamento de Astronom\'{\i}a y Astrof\'{\i}sica, Universitat de Val\`encia, Avinguda Vicent Andrés Estellés 19, 46100, Burjassot (Val\`encia), Spain}
\author{Miquel Miravet-Tenés\orcidlink{0000-0002-8766-1156}}
\affiliation{Mathematical Sciences and STAG Research Centre, University of Southampton, Southampton SO17 1BJ, UK}
\affiliation{Departamento de Astronom\'{\i}a y Astrof\'{\i}sica, Universitat de Val\`encia, Avinguda Vicent Andrés Estellés 19, 46100, Burjassot (Val\`encia), Spain}
\author{Milton Ruiz\orcidlink{0000-0002-7532-4144}}
\affiliation{Departamento de Astronom\'{\i}a y Astrof\'{\i}sica, Universitat de Val\`encia, Avinguda Vicent Andrés Estellés 19, 46100, Burjassot (Val\`encia), Spain}
\author{Antonios Tsokaros\orcidlink{0000-0003-2242-8924}}
\affiliation{Department of Physics, University of Illinois at Urbana-Champaign, Urbana, IL 61801, USA}   
\affiliation{National Center for Supercomputing Applications, University of Illinois at Urbana-Champaign, Urbana, IL 61801, USA}
\affiliation{Research Center for Astronomy and Applied Mathematics, Academy of Athens, Athens 11527, Greece}
\author{Stuart L. Shapiro\orcidlink{0000-0002-3263-7386}}
\affiliation{Department of Physics, University of Illinois at Urbana-Champaign, Urbana, IL 61801, USA}

\date{\today}

\begin{abstract}
  We perform general relativistic magnetohydrodynamic (GRMHD) simulations of binary neutron star (BNS) mergers with four distinct magnetic field topologies: (i) a dipole pulsar-like configuration, (ii) a mixed linear superposition of poloidal and toroidal components inside the star, and (iii-iv) two topologies featuring a smooth transition from a confined mixed core to a pulsar-like structure at radii $0.95\,R_{\rm NS}$ and $0.5\,R_{\rm NS}$, with $R_{\rm NS}$ the radius of the star. The latter topologies are explored in BNS merger studies for the first time. We evolve systems with two equations of state (EoS), SLy and WFF1, with ADM masses 2.7 and 2.6, respectively, and include an additional lower-mass SLy binary to probe the behavior of long-lived remnants. We perform an extensive analysis of the emission properties of the systems, both electromagnetic and gravitational waves, and of the properties of the remnants, namely their frequency modes, density eigenfunctions, rotation, temperature, and convective stability. We report three key results: (1) for the first time, we assess the convective stability of magnetized remnants, extending previous unmagnetized analyses; (2) we identify a clear secondary peak in the gravitational-wave spectrum of pulsar-like configurations, consistent with the nonlinear coupling of the $m=0$ and $m=2$ modes, which is absent in other topologies; and (3) the magnetic field topology strongly influences the gravitational wave emission properties to the extent that nearby ($<50\,{\rm Mpc}$) events could allow one to observationally distinguish between different field structures with future gravitational-wave detectors. Across all models, we obtain luminosities compatible with short gamma-ray bursts (sGRBs), with purely poloidal configurations being the most efficient in driving possible relativistic jets.
\end{abstract}
\maketitle

\section{Introduction}
\label{sec:intro}
The coincident detection of gravitational waves (GWs) from GW170817, and their electromagnetic (EM) counterparts across the spectrum, marked the beginning of multimessenger astronomy. This event, along with GRB170817A, confirmed that binary neutron star (BNS) mergers can be the progenitors of the central engine that powers short gamma-ray bursts (sGRBs)~\cite{LIGOScientific:2017vwq, LIGOScientific:2017ync, LIGOScientific:2017zic, LIGOScientific:2017pwl, LIGOScientific:2020aai, Pozanenko:2019lwh}. Notice that, although the progenitor of GW170817 has been officially classified as a BNS, since the estimated individual masses are consistent with the observed  mass distribution of neutron stars (NSs) and black holes (BHs) in binaries~\cite{Miller:2021qha,Demorest:2010bx,Ascenzi:2024wws},  observations cannot rule out the possibility of a neutron star-black hole (NSBH) system. However, these values fall below the minimum BH mass limit determined from X-ray binary observations and stellar collapse models~\cite{Orosz:2004ac}. While mechanisms for the formation of BHs with smaller masses have been proposed~\cite{Yang:2017gfb}, observational evidence for BHs below $3M_\odot$ is sparse, supporting the classification of GW170817 as a BNS merger.

This single multimessenger event has been used to impose tight constraints on the equation of state (EoS) at supranuclear densities~\cite{Margalit:2017dij,Shibata:2017xdx,Ruiz:2017due,Most:2018hfd,Rezzolla:2017aly}. It provided evidence of ejecta masses of approximately $0.01M_\odot-0.05M_\odot$ with velocities of $0.1c-0.3c$, consistent with the production of elements via the rapid neutron-capture process (r-process) needed to explain Milky Way abundances~\cite{Nicholl:2017ahq,Cowperthwaite:2017dyu,Arcavi:2017xiz}. Additionally, GW170817 offered an independent measure of the Hubble constant~\cite{LIGOScientific:2017adf}. This event underscored the importance of comparing multimessenger observations with theoretical models to understand extreme matter physics, highlighting the necessity of numerical modeling due to the complexity of the underlying phenomena.

The first numerical studies of BNSs began in the late 20th century, primarily focusing on understanding the basic dynamics of BNS mergers, while largely excluding the complexities introduced by magnetic fields or any microphysical processes \cite{Clark:1977,Wilson:1979,Shapiro:1980,Gilden:1984}. Early simulations included some degrees of relativistic gravity and simple hydrodynamic models \cite{Lai:1994jf,Rasio:1999ku, Rosswog:2000nj}. With advancements in computational capabilities and numerical methods, the 2000s saw the introduction of more sophisticated simulations incorporating full general relativity (GR) \cite{Shibata:2002jb, Duez:2002bn} and magnetohydrodynamics (MHD) \cite{Anderson:2008zp}. These improvements allowed the study of increasingly complex initial conditions \cite{Marronetti:2003gk, Baumgarte:2009fw, Tichy:2011gw, Tichy:2012rp, East:2012zn, Tsatsin:2013jca, Tacik:2015tja}, more realistic EoS \cite{Shibata:2005ss}, and higher numerical resolution~\cite{Kiuchi:2014hja, Kiuchi:2015sga}.
Despite the progress made in BNS merger simulations, there remains a need to enhance the modeling of magnetic reconnection, integrate more detailed microphysical processes, and deepen our understanding of the role of magnetic instabilities, and in particular, the role of the magnetic turbulence in the stability properties of the BNS remnant. Understanding these instabilities is crucial for interpreting the expected multimessenger signals from BNS systems, and may offer new insights into the mechanisms behind the launching of relativistic jets. However, magnetic fields significantly increase the complexity of simulations and introduce new computational challenges. To accurately capture magnetic instabilities, simulations with high numerical resolution are required. This has often rendered studies of magnetic instabilities unfeasible, due to the high computational costs of such simulations (see~\cite{Kiuchi:2024lpx} for a review). 
To overcome the need for high resolution, alternative approaches, explored in~\cite{Giacomazzo:2014qba,Aguilera-Miret:2020dhz,Aguilera-Miret:2021fre,Palenzuela:2021gdo,Palenzuela:2022kqk,Aguilera-Miret:2023qih, Miravet-Tenes:2025zkt}, have been proposed. They involve using large-eddy simulations and subgrid-scale models. These methods compensate for limited resolution by introducing additional terms in the evolution equations to approximate unresolved subgrid dynamics. A key advantage is that they allow for the use of weaker magnetic fields, closer to those observed in pulsars, and are considered to be more effective in modeling the small-scale, quasi-isotropic turbulent fields generated by the Kelvin-Helmholtz instability  (KHI)~\cite{Aguilera-Miret:2021fre}. These approaches are potentially preferred over artificially strong initial poloidal fields, which, as argued in~\cite{Aguilera-Miret:2023qih}, may produce unrealistic results. However, a notable drawback is that the outcomes may be highly sensitive to the choice of subgrid model and its parameterization, raising concerns about whether these models accurately reflect the true physical processes. Another complementary strategy is to employ approximate gravity treatments, such as the conformally flat condition, to reduce computational cost in the post-merger regime \cite{Jiang:2025ijp,Ng:2023yyg}.

Our GRMHD numerical simulations in~\cite{Ruiz:2016rai,Ruiz:2017inq,Bamber:2024kfb} with numerical resolutions ranging from $\Delta x\sim 90-220\rm\,m$ have shown that these mergers can produce a collimated, mildly relativistic outflow confined by a helical magnetic field emanating from the poles of a BH  surrounded by a magnetized accretion disk formed after the collapse of the hypermassive neutron star (HMNS) remnant.  This outflow, which is driven by the Blandford-Znajek (BZ) mechanism \cite{Blandford:1977ds}, was identified as an incipient jet. We used irrotational NSs of equal mass and a polytropic EoS with a polytropic index $\Gamma = 2$. Similar results were found in NSBH mergers~\cite{Paschalidis:2014qra,Ruiz:2018wah}. Follow-up studies included variations such as initial star spins~\cite{Ruiz:2019ezy}, different magnetic dipole orientations~\cite{Ruiz:2020via}, realistic piecewise polytropic EoSs (SLy and H4)~\cite{Ruiz:2021qmm}, and an M1 neutrino transport scheme~\cite{,Sun:2022vri}.
Ciolfi et al.~\cite{Ciolfi:2020hgg,Ciolfi:2019fie} conducted BNS merger simulations with numerical resolutions $\Delta x\sim 177-220\,\rm m$ resulting in a long-lived supramassive neutron star (SMNS) remnant, finding a collimated outflow but with heavy baryon pollution that prevents the formation of an outflow compatible with sGRBs. By contrast, Mösta et al.~\cite{Mosta:2020hlh} employed numerical resolutions  $\Delta x~\sim~ 55-220\,\rm m$ to perform pure hydrodynamic simulations of BNSs. The post-merger remnant was endowed with a poloidal magnetic field and reported the formation of a mildly relativistic jet, with neutrino cooling reducing baryon pollution and increasing the Lorentz factor, though still below sGRB levels. Most recently, Kiuchi et al.~\cite{Kiuchi:2023obe} performed very high-resolution simulations with $\Delta x\sim 12\,\rm m$  and noted the formation of a mildly relativistic jet, with potential for a higher Lorentz factor if efficient Poynting flux conversion occurs, despite severe baryon loading. Aguilera-Miret et al.~\cite{Aguilera-Miret:2024cor} using a resolution of $\Delta x\sim 120\,\rm m$ with a subgrid model observed a helical magnetic field  but found no other evidence of jet launching.

Most previous numerical studies assumed that the stars are initially endowed with dipole magnetic field, either confined entirely within the star's interior or extending from the interior to the exterior, similar to the standard picture for pulsars' exterior magnetic fields.

However, the interior magnetic field topology of NSs remains unknown. Observations of pulsar J0030+0451 with NASA’s Neutron star Interior Composition Explorer (NICER) have revealed unexpected hotspot locations, with up to three hotspots detected in the southern hemisphere, instead of the expected one at each pole~\cite{Miller:2019cac, Riley:2019yda, Dittmann:2024mbo}. These observations suggest the presence of a more complex magnetic field topology.
Moreover, both Newtonian and GRMHD simulations have shown that purely toroidal and purely poloidal magnetic field configurations are unstable~\cite{Lander_2010a, Lander_2010b, Ciolfi:2011xa, Ciolfi:2012en, Kiuchi:2008ss, Kiuchi:2011yt, Lasky:2011un, Tsokaros:2021pkh, Pinas:2025bpq}, consistent with analytical predictions \cite{10.1093/mnras/163.1.77, 10.1093/mnras/161.4.365, 10.1093/mnras/162.4.339}. 

NICER’s observations and the stability requirements for NS magnetic fields strongly advocate for the adoption of more sophisticated models to characterize their topology. For example, it has been suggested that NSs are born with both poloidal and toroidal fields exceeding $10^{14}\,\rm G$, but the poloidal component decays by a factor of $10$ over~$10^6\,\rm yr$~\cite{2012ApJ...754...27R}. As a first step, one can consider a superposition of toroidal and poloidal magnetic fields as the most general configuration. 

Multiple equilibrium solutions comprising mixed-field configurations have been proposed~\cite{10.1111/j.1365-2966.2008.12966.x}. Among these, the twisted-torus is a promising candidate for modeling magnetic fields in NS interiors~\cite{Ciolfi:2014jha}. In this configuration, the poloidal component of the field extends to the exterior of the star, while the toroidal one is confined within a torus inside the star, ensuring the continuity and smoothness of field lines at the stellar surface. However,  numerical simulations of isolated self-consistent rotating NSs with various twisted-torus configurations showed that stability depends strongly on the mixture of the different EM field topologies, with some configurations being more unstable than the others \cite{Tsokaros:2021pkh}.
Ongoing efforts have been devoted to the development of realistic NS magnetic field models~\cite{Gourgouliatos:2016fnl,Bucciantini:2015rre,Ciolfi:2009bv,Igoshev:2021ewx,Uryu2014,Uryu2019,Uryu2023}, with recent studies looking at the impact of these topologies on jet launching, the emitted gravitational wave signal, the global amplification of the magnetic field, and the nature of the ejected material \cite{Aguilera-Miret:2024cor,Bamber:2024qzi,Cook:2025frw,Gutierrez:2025gkx}.

Driven by the anticipated multimessenger observations of BNSs, we probe here the impact of distinct magnetic field topologies on the stability properties of the remnant, on the jet launching,  and more generally on both the EM and GW signals. We consider four different topologies: (i) a dipole poloidal pulsar-like configuration, (ii) a mixed linear superposition of poloidal and toroidal components inside the star, and (iii-iv) two topologies with a smooth transition from a confined mixed core to a pulsar-like structure at radii $0.95R_{\rm NS}$ and $0.5R_{\rm NS}$ (with $R_{\rm NS}$ the equatorial radius of the neutron star), ensuring physical plausibility and continuity across the stellar interior and magnetosphere. All configurations have exclusively poloidal fields outside the star, recovering the typical pulsar picture. Together, these span a broad range of plausible interior topologies. We find that systems with pulsar-like magnetic fields consistently exhibit the largest magnetic energies. These cases are consistent with observed sGRBs and kilonovae phenomena, expected to originate as the aftermath of BNS mergers~\cite{Cowperthwaite:2017dyu}. In scenarios where the remnant is a NS whose lifetime exceeds our simulated time, the magnetic field topology modifies the temperature distribution, regions of convective stability, and the remnant's structure. Consistent with~\cite{Bamber:2024qzi,Tsokaros:2024wgb}, we find that the $f_2$ GW peak from magnetized BNSs is shifted compared to purely hydrodynamic cases, with the shift being highly dependent on the magnetic topology. We do not consider neutrino effects here, as our focus is on magnetic field topology. However, a more complete analysis should include them, as neutrinos may drive jet formation without magnetic fields~\cite{Kawaguchi:2024naa} and reduce baryon pollution~({see~e.g.~\cite{Sun:2022vri,Just:2015dba}). 

The remainder of the paper is organized as follows. In Sec.~\ref{sec:methods}, we summarize the numerical methods to solve the GRMHD equations, the initial data used to model the BNSs, and the grid structure, along with a suite of diagnostics used to verify the reliability of our numerical calculations. Our main results are reported in Sec.~\ref{sec:results}.  Finally, we summarize our results and conclude in Sec.~\ref{sec:conclusions}. Additional comments on the magnetic power spectra and convective stability of the BNS remnants are provided in Appendices \ref{appendix:magnetic-power-spectrum} and \ref{appendix:convective-stability}, respectively. Throughout the paper, we adopt geometrized units ($G = c = 1$) except where otherwise explicitly stated. Greek indices denote all four spacetime dimensions, while Latin indices imply spatial components only.

\section{Methods}
\label{sec:methods}
The numerical methods employed to evolve our BNS models are the same as those in \cite{Etienne:2011ea,Ruiz:2021gsv}. In this section, we briefly introduce our notation and summarize our numerical schemes, along with the initial data. For more detailed information, we refer the reader to~\cite{Etienne:2011ea, Ruiz:2021gsv} and the references therein.

\subsection{Formulation and numerical scheme}
To solve the Einstein equations coupled with the equations of ideal MHD in a dynamic, curved spacetime, we utilize the {\tt Illinois GRMHD} code.  This code is the basis for its publicly available counterpart embedded in the {\tt Einstein Toolkit}~\cite{Etienne:2010ui, Werneck:2022exo} and is integrated into the {\tt Cactus} computational framework~\cite{CactusConfigs}. The  {\tt Illinois GRMHD} code employs {\tt Carpet} for moving mesh refinement~\cite{Carpet, carpetweb}, and it has been extensively tested and applied in various GRMHD contexts, including isolated BHs with magnetized accretion disks and BNS mergers undergoing delayed collapse~(for a review, see~\cite{Etienne:2011ea, Ruiz:2021gsv}). It solves the Baumgarte–Shapiro–Shibata–Nakamura (BSSN) formulation~\cite{Shibata:1995we, Baumgarte:1998te} to evolve the spacetime metric. This is coupled with the puncture gauge conditions, specifically the ``1+log" slicing for the lapse function and the hyperbolic gamma-driver condition for the shift cast in first order form (see Eq. (2)-(4) in~\cite{Etienne:2007jg}), using fourth order centered spatial differencing, except on shift advection terms, where a fourth order upwind differencing is used.  To mitigate high-frequency modes that may arise during the evolution, we apply fifth-order Kreiss-Oliger dissipation to all evolved variables, with a dissipation strength of 0.18.

The ideal MHD equations are solved in a flux conservative formulation, via high-resolution shock-capturing (HRSC) schemes
(see Eqs. (27)-(29)~in~\cite{Etienne:2010ui}), that employs the piecewise parabolic method (PPM)~\cite{COLELLA1984174} coupled to the Harten, Lax, and van Leer (HLL) approximate Riemann solver~\cite{doi:10.1137/1025002}. As a standard procedure to avoid the breakdown of these schemes we impose a tenuous atmosphere with rest-mass density $\rho_{\rm atm}=10^{-10}\rho_{0,\rm max}$, where $\rho_{0,\rm max}$ is the initial maximum value of the rest-mass density of the system.

Instead of evolving the magnetic field, we evolve the vector potential $A^\mu$ by integrating the magnetic induction equation 
which ensures that the ``no monopoles'' constraint is satisfied. We adopt the generalized Lorenz gauge, described 
in~\cite{Farris:2012ux}, to avoid the build-up of spurious magnetic fields.  Finally, we impose ``outflow'' boundary 
conditions to the primitive variables and the vector potential~\cite{Etienne:2007jg}. The time integration is performed 
with the method of lines using a fourth-order accurate Runge-Kutta method, with a Courant-Friedrichs-Lewy (CFL) factor of 
$0.45$.

\subsection{Initial data}
We consider equal-mass BNS configurations in quasi-equilibrium circular orbits that merge to form either a stable remnant, below the supramassive limit, or a transient HMNS that will undergo delayed collapse into a BH. The binaries consist of two identical irrotational NSs, modeled by a piecewise polytropic representation of the Skyrme-Lyon (SLy)~\cite{1998NuPhA.635..231C} and  the WFF1~\cite{Wiringa:1988tp}  nuclear EoSs. The initial data are computed using the Parallel Compact Object CALculator ({\tt PCOCAL}) code \cite{Boukas:2023ckb,Tsokaros2015}, and the properties of the configurations considered are summarized in Table~\ref{table: initial data}.

We note that these two EoSs broadly align with the current observational constraints on NSs. For example, the maximum mass configuration of an isolated star predicted by SLy is $M_{\rm sph}^{\rm max}=2.05M_\odot$, while that predicted by WFF1 is $M_{\rm sph}^{\rm max}=2.12M_\odot$. Both are consistent with: i) $M_{\rm sph}^{\rm max}>2.072^{+0.067}_{-0.066}M_\odot$ from the NICER and XMM analysis of PSR J0740+6620~\cite{Riley:2021pdl}; ii) $M_{\rm sph}^{\rm max}>2.01^{+0.017}_{-0.017}M_\odot$ from the NANOGrav analysis of PSR J1614-2230~\cite{Fonseca:2016tux}; iii) $M_{\rm sph}^{\rm max}>2.01^{+0.14}_{-0.14}M_\odot$ from the pulsar timing analysis of PSR 0348+0432~\cite{Antoniadis:2013pzd}; and iv) $M_{\rm sph}^{\rm max}>2.14^{+0.20}_{-0.18}M_\odot$ from the NANOGrav and the Green Bank Telescope~\cite{NANOGrav:2019jur}.
We also note that SLy predicts that a star with a mass of $1.4M_\odot$ has a radius of $R= 11.46\,\rm km$, consistent with the value $R=11.94^{+0.76}_{-0.87}\rm\, km$ obtained by a combined analysis of X-ray and GW measurements of PSR J0740+6620~\cite{Pang:2021jta}. Such a star modeled with WFF1 has a radius of $R = 10.23 \,\rm km$, which is slightly below the constraint imposed by~\cite{Pang:2021jta}. Consistent with this, the combined analysis of the LIGO/Virgo/KAGRA (LVK) collaboration of the progenitors of GW170817 with the radio-timing observations of the pulsar J0348+0432~\cite{LIGOScientific:2018cki,Antoniadis:2013pzd} constrains the radius of a NS with mass in the range $1.16M_\odot-1.6M_\odot$ to be $11.9^{+1.4}_{-1.4}\,\rm km$ at the $90\%$ credible level. However, the NICER analysis of PSR J0030+0451~\cite{Miller:2019cac} constrains the radius of a NS with mass of $1.44^{+0.15}_{-0.14}M_\odot$ to be $R= 13.02^{+1.24}_{-1.06}\, \rm km$. The SLy and WFF1 EoSs predict that a $1.44M_\odot$ star has a radius of $11.45\,\rm km$ and $10.24 \,\rm km$, respectively.
Therefore, SLy is slightly below the NICER constraint, while WFF1 deviates by at least $14\%$. Consequently, NICER analysis favors stiffer EoSs. Furthermore, the LVK analysis of GW170817 predicts that the tidal deformability of a $1.4\,M_\odot$ NS is $\Lambda_{1.4}= 190^{+390}_{-120}$  at the $90\%$ credible level \cite{LIGOScientific:2018cki}. For such a star, the tidal deformability values are $\Lambda_{1.4}=306.4$ for SLy and $\Lambda_{1.4}=157.33$ for WFF1. Hence, LVK analysis favors softer EoSs, such as SLy and WFF1~\cite{Tsokaros:2020hli}. On the other hand, the analysis of GW190814 \cite{LIGOScientific:2020zkf} suggests a mass of $2.59^{+0.08}_{-0.09}M_\odot$ for the secondary object, which favors stiffer EoS. Note, however, that the constraints imposed by event GW190814 are more uncertain as there is no confirmation that the secondary object is a neutron star.

\begin{table}
    \caption{Summary of the initial properties of the BNS configurations. We list the EoS employed to model the NSs, the equatorial coordinate radius $R_{\rm NS}$ toward the companion, the compactness $\mathcal{C}$, the tidal deformability $\Lambda = (2/3) k_2 \mathcal{C}^{-5}$ (where $k_2$ is the second Love number), the rest-mass of the binary $M_0$, the ADM mass $M$, and the angular velocity $\Omega$ for an initial binary coordinate separation of 
    $\sim 45 \,{\rm km}$.} 
    \label{table: initial data}
    \centering 
    \begin{tabular}{cccccccc}
        \toprule[0.5pt]
        \rule{0pt}{10pt}EoS       & $R_{\rm NS}[{\rm km}]$ & $\mathcal{C}$ & $\Lambda$ & $M_0[\Msol]$ & $M[\Msol]$\footnote{We note that the threshold gravitational mass at infinite separation for prompt collapse is $2.83M_{\odot}$ and $2.68M_{\odot}$ for SLy and WFF1, respectively~\cite{Bauswein:2020aag}.} & $M\Omega$ \\ \midrule[0.5pt]
          \rule{0pt}{10pt}SLy     & 9.05            &  0.17         & 353       & 3.04         & 2.70       & 0.030 \\
         \rule{0pt}{10pt}SLy     & 9.25            &  0.16         & 498       & 2.83         & 2.54       &  0.029 \\
         \rule{0pt}{10pt}WFF1     & 7.86            & 0.19         & 232       & 2.95         & 2.60       & 0.037     \\
         \bottomrule[0.5pt]
    \end{tabular}
\end{table}

\subsection{Grid structure}
\label{Sec:grid}
Details on the grid structure are given in Table~\ref{table: grid setup}. It consists of two sets of nested refinement boxes, each centered on one of the NSs. Each set contains nine boxes, with sizes and resolutions differing by factors of two. When two boxes overlap, they are merged into a single box centered on the center of mass of the binary system. The finest box surrounding each NS has a half-length on the side of approximately $1.25$ times the initial equatorial radius $R_{\rm NS}$ of the NS. This configuration allows us to resolve the NS equatorial radius with $\gtrsim 90$ grid points similar to our previous studies in~\cite{Bamber:2024kfb,Bamber:2024qzi}. Since we do not expect reflection symmetry across the equatorial plane to strongly impact the outcome of our simulations \cite{Ruiz:2020via}, we impose it to reduce computational costs.

\begin{table}[H]
    \centering 
      \caption{Grid structure for models listed in Table \ref{table: initial data}. The computational mesh consists of two sets of 9 nested refinement boxes. 
      The finest boxes are centered on each star and have a half-length of $\sim 1.25R_{\rm NS}$, where $R_{\rm NS}$ is the initial equatorial coordinate radius toward the companion. 
      The number of grid points covering the equatorial radius of the NS is denoted by $N_{\rm NS}$. In all cases, we impose reflection (equatorial) symmetry about the orbital plane. For the lower-mass SLy case, we include the ADM mass in the EoS name to distinguish it from the higher-mass case.}
    \label{table: grid setup}
    \begin{tabular}{ccccccc}
        \toprule[0.5pt]
        \rule{0pt}{10pt}EoS & Grid hierarchy$^{\dag}$ & Max. resolution & $N_{\rm NS}$ \\ \midrule[0.5pt]
          \rule{0pt}{10pt}SLy & 3024.03\,km/2$^{n-1}$ & 95 m & 91 \\
         \rule{0pt}{10pt}SLy2.54 & 5748\,km/2$^{n-1}$ & 90 m & 102 \\
         \rule{0pt}{10pt}WFF1 & 2646.03\,km/2$^{n-1}$ & 86 m & 91 \\
         \bottomrule[0.5pt]
    \end{tabular}
    \begin{flushleft}
  $^{\dag}$ The spatial extent of the simulation box for each refinement level, where $n$ denotes the level number.
\end{flushleft}
\end{table}

\begin{figure*}
\begin{turn}{90}
\hspace{1.7cm}\bf Pulsar-like
\end{turn}
\includegraphics[scale=0.129]{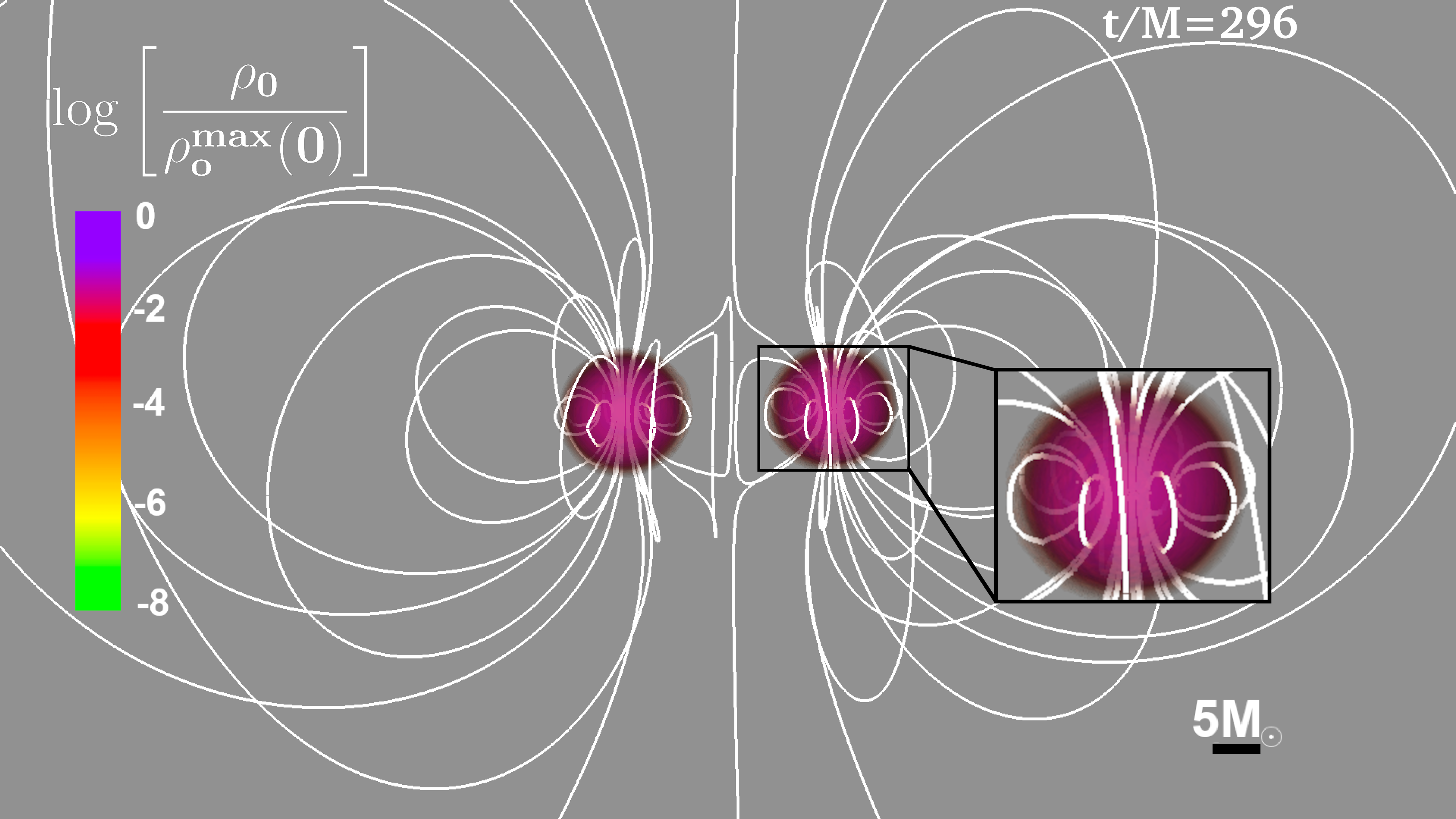}
\includegraphics[scale=0.129]{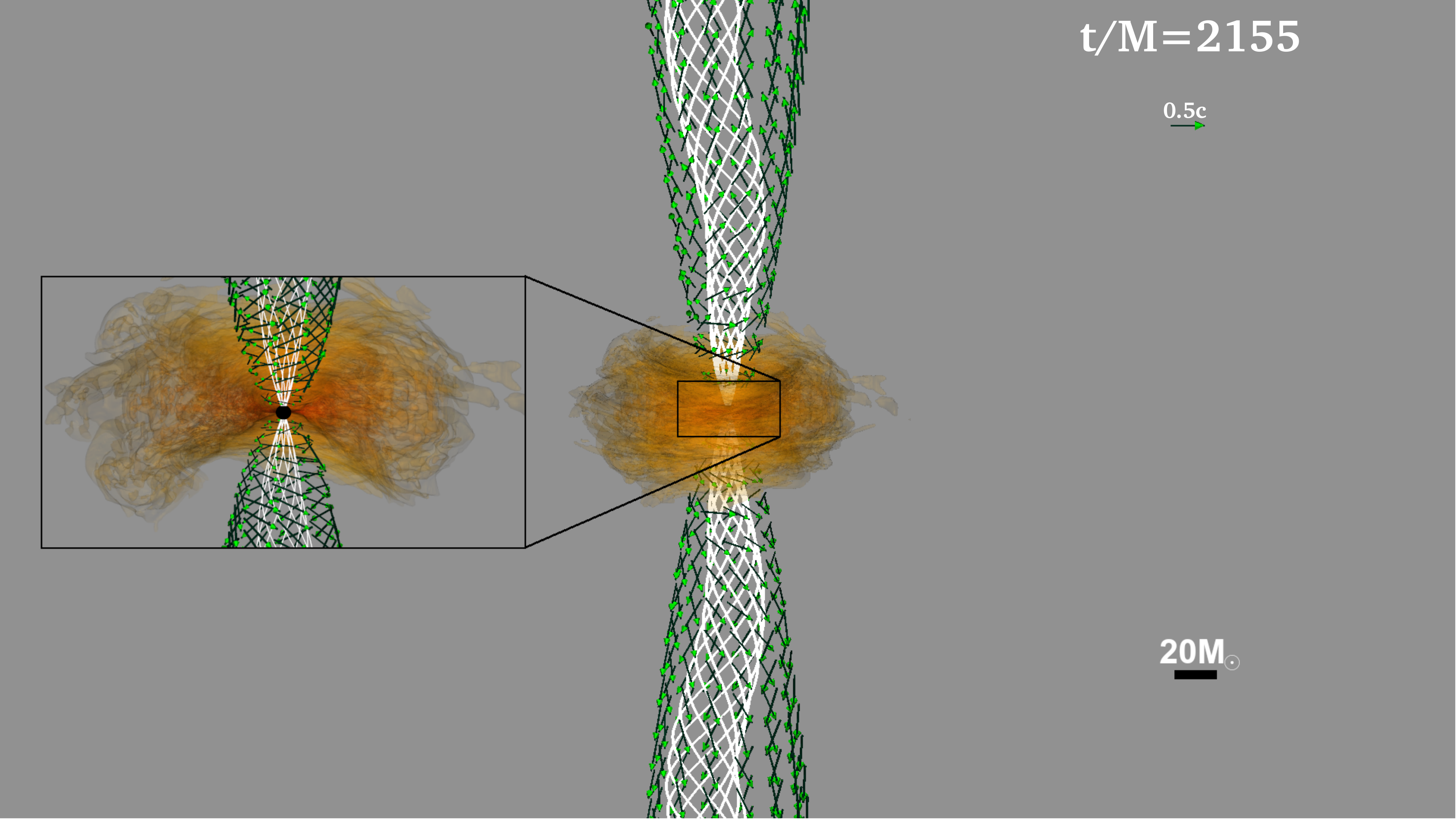}
\begin{turn}{90}
\hspace{1.1cm}\bf Pulsar-like + toroidal
\end{turn}
\includegraphics[scale=0.129]{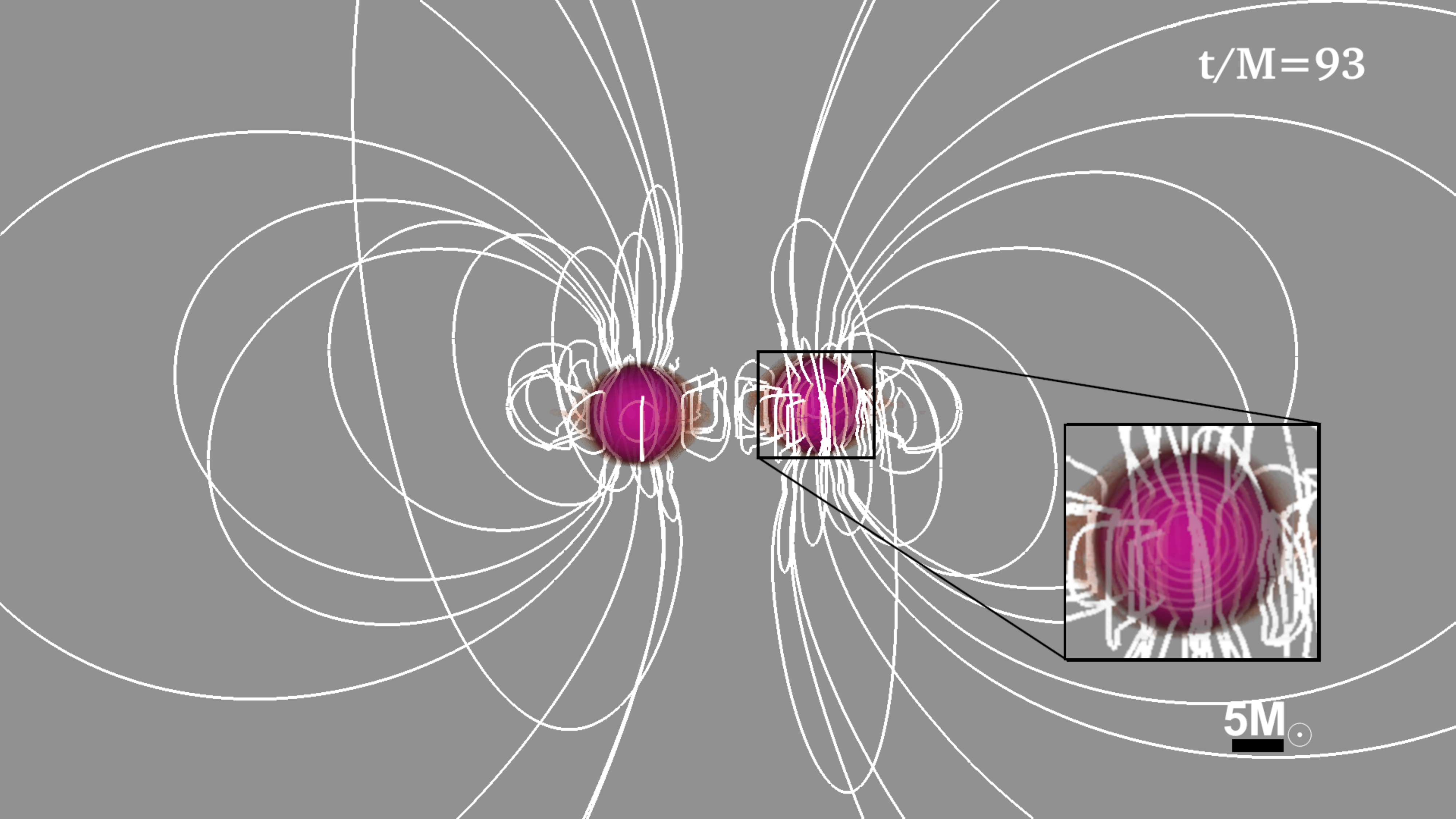}
\includegraphics[scale=0.129]{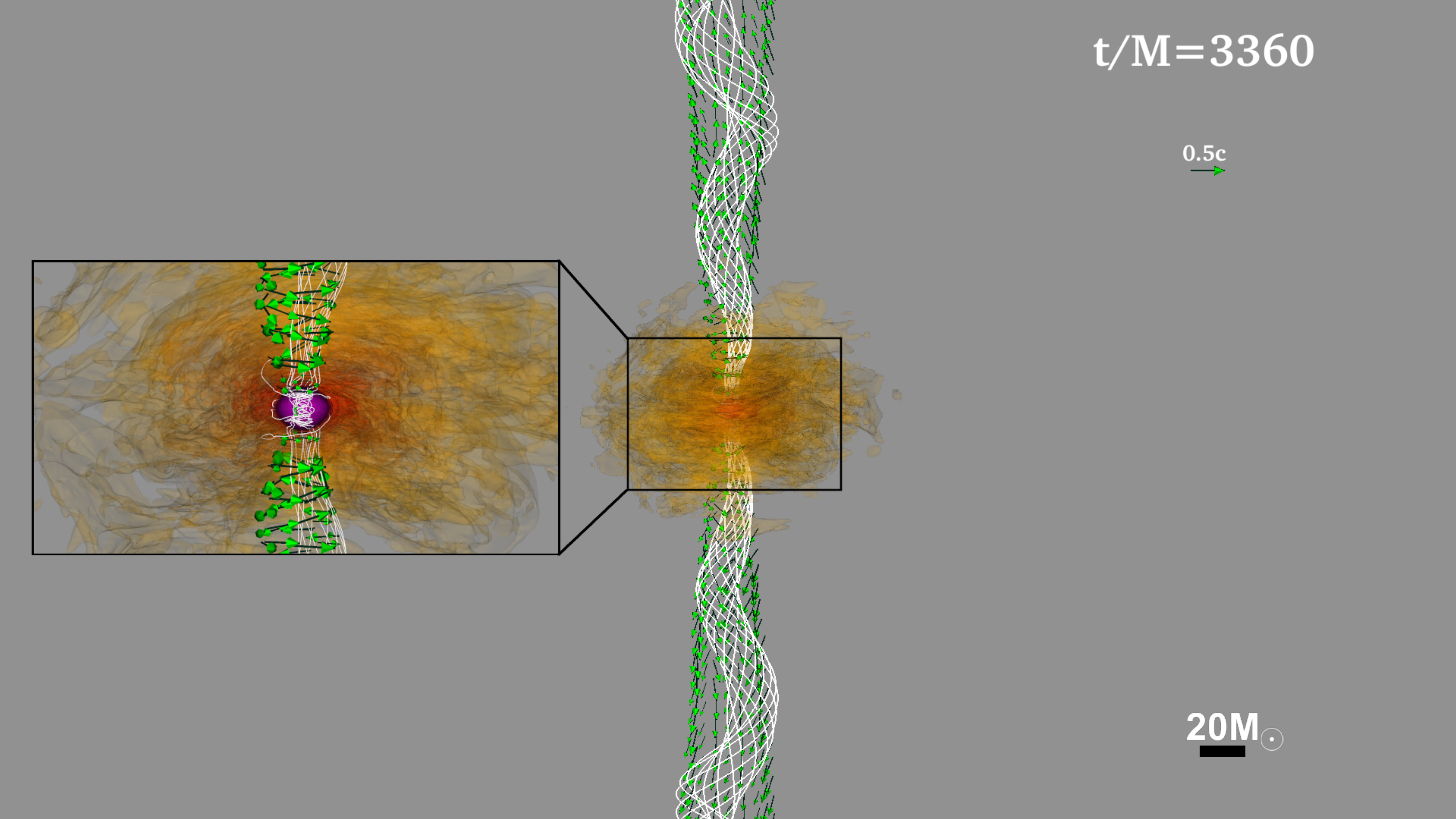}
\caption{Volume renderings of the rest-mass density, normalized to its initial maximum value (logarithmic scale), at the moment of magnetic field insertion (left) and after the BNS remnant reaches quasiequilibrium (right). Top panel shows the SLy BNS with a pure poloidal (pulsar-like) magnetic field, while the bottom panel depicts the WFF1 BNS with a mixed magnetic field inside the star ($\rone=0.95\,R_{\rm NS}$, see~Sec.~\ref{sec:magnetic-fields}),  smoothly transitioning to an external dipole  field. Insets highlight the initial magnetic field configuration inside the star (left), and the helical magnetic structure after an incipient jet is launched from the pole of the the binary's remnant (right top panel BH remnant and right bottom NS remnant). White lines represent magnetic field lines, arrows indicate plasma velocities, and the black sphere marks the BH apparent horizon. Here, $M=2.7\Msol$ for the top panels and $M=2.6\Msol$ for the bottom panels, with $\Msol=4.9\times10^{-3}\,{\rm ms} = 1.4\,{\rm km}$.}
\label{fig:visit-3dplot-SLy-case1}
\end{figure*}

\subsection{Magnetic field models}
\label{sec:magnetic-fields}
To mitigate the effects of magnetic winding during the inspiral phase, we evolve the configurations in
Table~\ref{table: initial data} with no magnetic fields to approximately two orbits ($\lesssim 3\,{\rm ms}$, 
roughly smaller than {the} Alfvén {timescale}; see below) 
before merger, except for the lower-mass SLy configuration.\footnote{Note that 
the lower-mass SLy configuration (which we denote by SLy2.54) initially threaded by a pulsar-like poloidal magnetic field (see Eq.~\eqref{eq:poloidal}), and reported previously in~\cite{Bamber:2024kfb}, is considered in our study to probe the effect of magnetic fields in a long-lived remnant.}
At that point, the NSs are threaded by a dynamically weak magnetic field generated by the vector potential
\begin{equation}
    \mathbf{A} = (1-a_T)\,\mathbf{\AP} + a_T\,\mathbf{\AT} \, ,
    \label{eq:Vec_Pot}
\end{equation}
where $\mathbf{\AP}$ generates a poloidal magnetic field component given by
\begin{equation} 
\label{eq:poloidal}
    \mathbf{\AP} = \frac{\pi\,\varpi^2\,I_P\,r_P^2}{(r_P^2+r^2)^{3/2}}\left[1+\frac{15\,r_P^2(r_P^2+\varpi^2)}{8\,(r_P^2+r^2)^2}\right] \bf{\hat{e}_\phi}\, ,
\end{equation}
that approximately corresponds to that generated by a current loop with radius $r_P=4.3$ inside the star, where $I_P$ is the current, set to $\sim7.2\times10^{-5}$ and $\sim4.5\times10^{-3}$ for the SLy and WFF1 binaries, respectively, $r^2=\varpi^2+z^2$, with $\varpi^2=(x-x_{\rm NS})^2+(y-y_{\rm NS})^2$ and $(x_{\rm NS}, y_{\rm NS}, 0)$ is the coordinate location of the center of mass of the NS. The vector potential $\mathbf{\AT}$ generates a tilted toroidal magnetic field component and is given by

\begin{equation} \label{eq:toroidal}
    \mathbf{\AT} = 2\,I_T\,\log\left(\frac{2\,r_T}{r}\right)\max (P- P_{\rm cut},0)^{n_b}\, 
    \bf{\hat{e}_y} \, ,
\end{equation}

where $r_T$ and $I_T$ are the loop radius and the current that generates the toroidal field, analogous to $r_P$ and $I_P$, respectively. We set $r_T = 4.3$, and $I_T\sim 3.786$ and $\sim 1.576$, for the SLy and WFF1 binaries, respectively. The cutoff pressure parameter $P_{\rm cut}$, which we set to 0.1\,$P_{\rm max}$ (with $P_{\rm max}$ the maximum initial pressure), confines the toroidal component to reside within $P>P_{\rm cut}$. 
Note that in contrast to~\cite{Bamber:2024qzi}, the toroidal component is chosen to be perpendicular to the orbital angular momentum, rather than aligned with it, hence being called ``tilted". This means that the vector potential generating this component lies within the orbital plane. This component generates a mixed magnetic field, with both poloidal and toroidal components. The parameter $n_b$ determines the degree of central condensation of the magnetic field and is set to 1. 

We consider the following magnetic field topologies: 
\begin{itemize}
    \item \textbf{Pulsar-like magnetic field (P)}: Following~\cite{Ruiz:2021gsv}, the stars are endowed with a purely poloidal magnetic field, 
    characterized by $a_T=0$ in Eq.~\eqref{eq:Vec_Pot} (see top-left panel in~Fig.~\ref{fig:visit-3dplot-SLy-case1}). This serves as our fiducial case.
    \item \textbf{Superposition of poloidal + toroidal magnetic field (SP)}: The stars are endowed with a mixed magnetic field configuration within their interior, 
    consisting of a linear superposition of toroidal and poloidal components, and in the exterior we impose a pulsar-like magnetic field. In this case, 
    we set $a_T=0.5$ in~Eq.~\eqref{eq:Vec_Pot}.
    \item \textbf{Tilted toroidal magnetic field in the NS + external pulsar-like field (T)}: The stars are endowed with a mixed magnetic field within their interior, which smoothly transitions to a pulsar-like magnetic field extending from the interior to the exterior (see bottom-left panel 
    in~Fig.~\ref{fig:visit-3dplot-SLy-case1}). To achieve a smooth transition between the interior and exterior magnetic field components, we set $a_T=\expfac$ as in~\cite{Sun:2017voo}, where $\rone$ and $p$ are free parameters that determine the degree of central condensation of the interior magnetic field component, with $r_1$ indicating where the transition occurs, and $p$ setting how fast this transition is. For this magnetic field configuration, we consider two cases: i) $\rone=0.95\,R_{\rm NS}$, and denote it {\bf{T}}$_{0.95}$, 
    which corresponds to an essentially mixed field confined in the star; and ii) $\rone=0.5\,R_{\rm NS}$,  where the mixed field is more confined to the star's core, which is denoted as {\bf{T}}$_{0.5}$. In both cases, we set $p=10$.
\end{itemize}
The only difference between configurations {\bf{SP}} and {\bf{T}} is the value of $a_T$, which in case {\bf{T}} ensures a smooth transition and also allows to confine the toroidal component (in the {\bf{SP}} configuration it is confined by the pressure, guaranteeing that there is no toroidal component outside the stars), whereas the confinement in configuration {\bf{T}} is determined by $r_1$. The equation for the vector potential is given in Eqs.~\eqref{eq:Vec_Pot},~\eqref{eq:poloidal}, and~\eqref{eq:toroidal}, with $a_T$ the one stated for each configuration.

To differentiate between the different cases, we labeled them as EoS$\_$magnetic-field-configuration. For the confined mixed cases,
an additional tag is used to indicate the confinement level of the tilted toroidal component (see Table~\ref{table: initial magnetic field}). Additionally, for the lower-mass SLy binary, we include a tag for its ADM mass, to distinguish it from the heavier configurations with the same EoS.
For example, SLy$\_$P denotes a BNS configuration modeled with the SLy EoS, ADM mass of $2.7\Msol$, and a purely poloidal magnetic field, while SLy$\_$T$_{0.5}$ represents the same binary but with a mixed magnetic field confined within the star, with $\rone=0.5\,R_{\rm NS}$~(see 
Sec.~\ref{sec:magnetic-fields}). Cases with no magnetic fields (hydrodynamical only) are denoted by the tag H.

To ensure a fair comparison of the evolution across the different magnetic field configurations, we choose the above free parameters such that the maximum magnetic-to-gas pressure ratio within the interior of the NSs is~$\beta^{-1}\equiv P_{\rm mag}/P=0.003125$ across all cases, as in~\cite{Ruiz:2016rai,Ruiz:2017inq}, which corresponds to a maximum magnetic field as measured by a normal observer of $B_{\rm max} \lesssim 10^{16.4}\, \rm G$ (see Table~\ref{table: initial magnetic field}).  This value is significantly higher than the surface magnetic field strengths of~$\sim 10^8-10^{12}\,\rm G$ typically observed in NSs within binary systems, as inferred from binary pulsar observations~\cite{Tauris:2017omb,Lorimer:2008se}. Following~\cite{Ruiz:2016rai,Ruiz:2017inq}, we choose this value to mimic the amplified field strengths expected from the exponential growth of magnetic instabilities, which has been observed in high resolution BNS simulations~\cite{Kiuchi:2017zzg,Palenzuela:2021gdo} (see Sec.~\ref{Sec:mag_instabilities} for a more detailed discussion on magnetic instabilities). 
We find that the magnetic power spectra of our BNS remnants are in broad agreement with higher-resolution and large-eddy simulations~\cite{Kiuchi:2015sga, Aguilera-Miret:2020dhz} (see Appendix~\ref{appendix:magnetic-power-spectrum}).}

We also consider a slightly weaker pulsar-like magnetic field configuration by setting $\beta^{-1}\equiv P_{\rm mag}/P=0.0023$ for the SLy2.54\_P case, as in~\cite{Bamber:2024kfb}. As shown below, the remnant in this case behaves similarly to those modeled with the WFF1 EoS (it does not collapse), allowing us to analyze the stability properties
of longer-lived NS remnants.  
\begin{table}
    \centering 
 \caption{Initial total magnetic field energy $\mathcal{M}$, and its toroidal ($\mathcal{M}_{\rm tor}$) and poloidal ($\mathcal{M}_{\rm pol}$) components, all in erg, maximum magnetic filed strength $B_{\rm{max}}$, and magnetic field strength at the pole $B_{\rm{pole}}$, both in G, for all BNS 
    configurations in Table~\ref{table: initial data} in which all cases have a maximum value of $\beta^{-1}\equiv 
    P_{\rm mag}/P=0.003125$ within the star, except for the  $\rm SLy2.54\_P$ configuration in which $\beta^{-1}=0.0023$.}
    \begin{tabular}{cccccc}
        \toprule[0.5pt]
        \rule{0pt}{10pt}Case             & $\mathcal{M}$[erg] & $\mathcal{M}_{\rm tor}$[erg] & $\mathcal{M}_{\rm pol}$[erg] & $B_{\rm{max}} [G]$ & $B_{\rm{pole}} [G]$ \\ \midrule[0.5pt]
          \rule{0pt}{10pt}SLy\_P         & $10^{50.66}$ & $10^{20.92}$ & $10^{50.66}$ & $10^{16.23}$ & $10^{14.05}$ \\
         \rule{0pt}{10pt}SLy2.54\_P      & $10^{50.61}$& $10^{20.87}$ &  $10^{50.61}$ &$10^{16.20}$  & $10^{15.96}$ \\         
         \rule{0pt}{10pt}SLy\_SP         & $10^{48.72}$& $10^{48.12}$ & $10^{48.60}$ &$10^{15.26}$ & $10^{12.21}$ \\
         \rule{0pt}{10pt}SLy\_T$_{0.95}$ & $10^{50.53}$& $10^{49.93}$ & $10^{50.41}$ &$10^{16.17}$  & $10^{13.44}$\\
         \rule{0pt}{10pt}SLy\_T$_{0.5}$  & $10^{49.96}$& $10^{49.36}$ & $10^{49.84}$ &$10^{15.88}$  & $10^{10.56}$\\ \hline
         \rule{0pt}{10pt}WFF1\_P         & $10^{50.82}$&  $10^{21.09}$& $10^{50.82}$&$10^{16.41}$  & $10^{14.32}$\\
         \rule{0pt}{10pt}WFF1\_SP        & $10^{48.86}$& $10^{48.26}$ & $10^{48.74}$ &$10^{15.43}$ &  $10^{13.30}$\\
         \rule{0pt}{10pt}WFF1\_T$_{0.95}$& $10^{50.39}$& $10^{49.79}$ & $10^{50.27}$ &$10^{16.19}$  & $10^{13.67}$\\
         \rule{0pt}{10pt}WFF1\_T$_{0.5}$ & $10^{50.05}$& $10^{49.45}$ & $10^{49.93}$ &$10^{16.02}$ & $10^{10.73}$\\ 
         \bottomrule[0.5pt]
    \end{tabular}
    \label{table: initial magnetic field}
\end{table}

\subsection{Magnetic instabilties}
\label{Sec:mag_instabilities}
The KHI acts on small scales through the small-scale turbulent dynamo mechanism~\cite{Schekochihin:2005rd,Chandrasekhar:1961}. 
The turbulent flow resulting from the shearing of fluid flowing in opposite directions stretches and folds the magnetic field lines. 
The stretching increases the length of the field lines, enhancing the magnetic field tension along the direction of stretching due to the 
frozen-in property of magnetic fields in a high-conductivity fluid. In highly turbulent regions, magnetic diffusion acts to dissipate the 
magnetic energy. Eventually, the magnetic field grows to a point where its back-reaction on the fluid flow becomes significant. This back-reaction 
modifies the turbulence, reducing the efficiency of the dynamo, reaching saturation. The growth rate of the KHI, $\sigma_{\rm{KHI}},$ is inversely proportional 
to the wavelength~\cite{Schekochihin:2005rd,Chandrasekhar:1961}
\begin{equation}\label{eq:KHI}
\sigma_{\rm KHI} \sim \frac{\Delta v}{\lambda}\,,
\end{equation}
where $\Delta v$ is the difference in velocity across the fluid layers that come into contact and move past each other. Eq.~\eqref{eq:KHI} 
suggests that the small-scale vortices grow faster than the large-scale vortices. However, in numerical simulations 
the wavelength is limited by the numerical resolution. Note that an initial magnetic field strength of $10^{11}\,\rm G$ 
requires a numerical resolution of at least $\sim 7\,\rm m$ to  properly capture the magnetic field amplification due 
to the KHI~\cite{Kiuchi:2023obe}. Given the limited computational resources available, we employ stronger initial fields to
mimic the magnetic field amplification due to the KHI. 
During the first $2\,{\rm ms}$ following merger we observe that, consistent with the values reported in~\cite{Bamber:2024kfb},  
the growth rate of the total magnetic energy is $\sim 1.18\,{\rm ms}^{-1}$ for  SLy binaries and $\sim 1.37 \,{\rm ms}^{-1}$ for 
those modeled with WFF1. The slightly higher  amplification  factor in the latter cases 
is likely due to a higher  speed difference across the shear surface.
As in~\cite{Kiuchi:2023obe, Bamber:2024kfb}, the measured growth rate is two orders of magnitude lower than the expected $\sigma_{\rm KHI}\sim10^2\,{\rm{ms}}^{-1}$ from perturbation theory.

After the BNS remnant settles into a massive central core wrapped by a low-density cloud of matter, the high velocity 
gradients that drive the KH instability weaken. Amplified magnetic fields then regulate the flow through angular momentum 
transport, suppressing the conditions necessary for the KHI to continue.
The magnetorotational instability (MRI) and magnetic winding emerge during the post-merger phase of BNS mergers,  enhancing magnetic turbulence and 
angular momentum transport \cite{Giacomazzo:2014qba, Palenzuela:2022kqk, Gutierrez:2025gkx, Cook:2025frw}.

The remnant becomes unstable to the MRI when the condition ${\partial\Omega^2}/{\partial\ln \varpi}<0$, with $\Omega$ the angular velocity, is satisfied~\cite{1991ApJ...376..214B,1998RvMP...70....1B}.
Two fluid elements in the remnant, connected by a magnetic field line, experience different rotational speeds. The inner element, closer to the stellar center, rotates faster than the outer one. This differential rotation stretches the magnetic field line, generating magnetic tension.
The faster-rotating inner element pulls on the magnetic field line, exerting a force on the slower-rotating outer element, thereby increasing its angular velocity. Conversely, the outer element exerts a retarding force on the inner element, decreasing its angular velocity. This exchange of angular momentum between the elements destabilizes the rotational equilibrium, causing the inner element to move inwards (losing angular momentum) and the outer element to move outwards (gaining angular momentum). This process of stretching the magnetic field lines and exchanging angular momentum intensifies, leading to an increase in magnetic tension and further amplification of the velocity perturbations. Consequently, this feedback loop results in the exponential growth of both the magnetic field and velocity perturbations, rapidly amplifying the strength of the magnetic field. The growth rate and the fastest-growing wavelength of the MRI read

\begin{align}
    \sigma_{\rm MRI} &= \frac{1}{2}\frac{\partial\Omega}{\partial\ln\varpi} \,, \\
    \lambda_{\rm MRI} &\approx \frac{2\pi v_A}{\Omega} \approx \frac{2\pi\sqrt{|b^P|^2/(b^2+\rho_0h)}}{\Omega} 
    \label{eq:MRI_grid}\\
   &\sim 18\,\rm cm\,
   \left(\tfrac{B}{10^{11}\rm G}\right)\,
   \left(\tfrac{\rho_0}{10^{14}\rm g\,cm^{-3}}\right)^{-1/2}\,
   \left(\tfrac{\Omega}{10^{3}\rm rad\,s^{-1}}\right)^{-1}\,,\nonumber
\end{align}
where $v_A$ is the Alfvén speed, $|b^P|=\sqrt{b^\mu b_\mu -(b_\mu(e_{\hat{\phi}})^\mu)^2}$, $b^\mu=B^\mu_{\rm co}/\sqrt{4\pi}$ where $B^\mu_{\rm co}$ is the magnetic field in the reference frame comoving with the fluid, and $(e_{\hat{\phi}})^\mu$ 
is the toroidal orthonormal vector comoving with the fluid, $\rho_0$ is the rest-mass density, and $h$ the specific enthalpy.

To determine if the MRI is adequately resolved in our simulations, two key conditions must be satisfied: i) the wavelength of 
the fastest-growing mode, $\lambda_{\rm MRI}$, has to be resolved by  at least $\simeq 10$ grid 
points; and ii) $\lambda_{\rm MRI}$ 
has to fit in the remnant~\cite{Sano:2003bf, Shiokawa:2011ih}. To check condition i), we compute the $\lambda_{\rm MRI}$-quality factor 
$Q_{\rm MRI}\equiv \lambda_{\rm MRI}/dx$, with $dx$ the local grid spacing, which measures the number of grid points per fastest 
growing MRI mode (see top panels of Fig.~\ref{fig:mri_c3}). We observe that the $Q\gtrsim 10$ criterion is satisfied in 
the binary remnant cases {\bf{P}}  and {\bf{T}} over a substantial portion of the remnant. However, in the {\bf{SP}} case the quality factor is lower. Note that fixing the maximum value of the magnetic-to-gas pressure ratio in the stellar interior to  $\beta^{-1}=0.003125$ induces different 
magnetic field strength depending on the topology of the magnetic field (see  Table~\ref{table: initial magnetic field}). The {\bf{SP}} configuration presents the lowest magnetic field strength, which helps explain the small values of $Q$ for that case, which is $\lesssim 5$ in the central core of the remnant, and around $\lesssim 10$ in the low density cloud of matter wrapping it since the MRI wavelength is proportional to $\rho_0^{-1/2}$ and therefore decreases in regions of higher density. Hence, we do not fully resolve the MRI for this configuration (see Table~\ref{tab:results}), and refrain from comparing any quantities, between {\bf{SP}} and other configurations, that could be influenced by whether the MRI is properly resolved or not.
To check condition ii), we plot the rest-mass density normalized to its maximum value along $\lambda_{\rm MRI}$ on the meridional plane.
In all cases, the fastest-growing mode $\lambda_{\rm MRI}$ fits within the remnant.

Magnetic winding occurs in differentially rotating objects when a poloidal magnetic field (with both radial and vertical components)
generates a linear growth of the toroidal magnetic field in a timescale of $\tau_{\rm wind}\sim \Delta\Omega^{-1}$,
where $\Delta\Omega$ is the difference in angular velocity across the star~\cite{Shibata:2021bbj}.
As the magnetic field lines are wound up and the toroidal component strengthens, magnetic tension develops, resisting 
differential rotation through magnetic braking. This process eventually drives the system toward uniform rotation 
on the Alfv\'en timescale~\cite{shibata2015numerical}
\begin{equation} \label{eq:Alfven-timescale}
    \tau_{\rm A} \sim 10^2\,{\rm ms} \left(\tfrac{\rho_0}{10^{15} 
    \rm g\,cm^{-3}}\right)^{1/2}\left(\tfrac{B}{10^{15}\rm G}\right)^{-1}\left(\tfrac{R}{10^{6}\rm cm}\right) \,.
\end{equation}
Note that this timescale is comparable to the lifetime of typical BNS remnants and may influence their subsequent
evolution~\cite{Kiuchi:2024lpx,Aguilera-Miret:2023qih}. As emphasized in~\cite{Kiuchi:2024lpx}, higher resolution
studies than those performed here are necessary to assess the effects of initially weak magnetic fields.

\subsection{Jets consistent with sGRBs}
Jet-like structures appear in a variety of natural phenomena across different scales in the universe, 
and are characterized as highly collimated streams of matter and energy ejected from a source \cite{Shapiro:2017cny}. 
In the context of stellar compact binary mergers, they are thought to lead to sGRBs, although the central 
engine is not yet completely understood. There is a general consensus about the possibility of such 
transients emerging from a BH surrounded by a magnetized accretion disk; however, there is no clear picture 
of whether sGRBs can emerge from NS binary remnants~\cite{Ruiz:2020zaz, Ruiz:2021gsv,Kiuchi:2024lpx}. However, an
increasing number of studies have been dedicated to exploring scenarios where GW170817-like events may 
leave behind a rapidly rotating NS~\cite{DuPont:2024sbz,Bamber:2024kfb,Kiuchi:2023obe}.

Numerical simulations have also shown jet-like structures emerging from stable NS 
remnants~\cite{Kiuchi:2024lpx,Bamber:2024kfb,Mosta:2020hlh,Palenzuela:2021gdo,Ciolfi:2019fie}, 
with recent studies focusing on how the $\alpha$-$\Omega$-dynamo mechanism can lead to large-scale magnetic 
fields in HMNSs, resulting in a Poynting flux-dominated relativistic outflow similar to that of the 
BZ mechanism for BHs~\cite{Kiuchi:2024lpx, Musolino:2024sju}.  
The formation of such jets from NSs hinges on dynamo processes influenced by microscopic parameters such as 
magnetic helicity.\footnote{A new formalism for the gauge-invariant computation of magnetic helicity transport in numerical relativity codes
has been presented in~\cite{Wu:2024dao}.} Some studies suggest that finite helicity can suppress  dynamo actions, hindering the 
organization of magnetic fields into large-scale structures ~\cite{Gruzinov:1994zz, 1995ApJ...449..739B, Blackman:2014kxa}. 
Additionally, inefficiencies in outflow acceleration for neutron star remnants have been observed in GRMHD 
simulations \cite{Kiuchi:2024lpx,Bamber:2024kfb,Mosta:2020hlh,Ciolfi:2019fie}. So, the question remains, 
are these incipient jets consistent with sGRBs?

To gain a better understanding of what systems can be the progenitors of the central engine that powers sGRBs, it 
is useful to study which jet properties are consistent with these EM signals. sGRBs are thought to originate from the internal 
and external shocks in a highly relativistic jet, with Lorentz factors of at least $\Gamma_{L}\gtrsim 20$, with 
typical values of~$\Gamma_{\rm L} \sim \mathcal{O}(10^2)$, and 90\% of the total $\gamma$-ray count is observed 
in $\lesssim 2\, {\rm s}$~\cite{Shappee:2017zly, Ghirlanda:2017opl, 1986ApJ...308L..43P, Granot:2005ye, 10.1111/j.1365-2966.2009.15863.x}.
The terminal Lorentz factor $\Gamma_\infty$ can be estimated near the base of the outflow by (see \cite{Bamber:2024kfb} for a more detailed discussion)
\begin{equation} \label{eq:Lorentz_infty}
    \Gamma_\infty \approx \frac{\mu}{2} \approx \frac{\sigma}{2} \approx \frac{b^2}{2\rho_0} = \frac{\rho_B}{\rho_0} \, ,
\end{equation}
where $\mu$ is the ratio between the energy flux and the rest-mass energy flux, $\sigma$ is the magnetization parameter, 
and $\rho_B = b^2/2 = B^2_{\rm co}/8\pi$ (where $b^\mu :=B^\mu_{\rm co}/\sqrt{4\pi}$ for 
comoving magnetic field $B^\mu_{\rm co}$) is the EM energy density. Whereas BH central engines demonstrate efficient acceleration 
capabilities (see~e.g.~\cite{Ruiz:2016rai,Hayashi:2024jwt,Just:2015dba}), the same is not as clear for stable NS remnants. 
Thus, understanding whether NS jets can emit sGRBs requires a rigorous assessment of baryon pollution, which is not within the scope of this work.

Despite lasting less than $2\,\rm s$, sGRBs release an extraordinary amount of energy, with typical isotropic-equivalent 
luminosities ranging from $\sim 10^{49} - 10^{52}\, {\rm erg\,s}^{-1}$~\cite{Li:2016pes, Beniamini:2020adb, Shapiro:2017cny}. 
Therefore, computing the isotropic-equivalent luminosities provides another method to assess whether the jets we observe 
in our simulations are capable of producing sGRBs. In addition, the duration of these transients correlates with the 
lifetime of the accretion disk (the jet's fuel) surrounding the remnant, which can be estimated by computing the accretion rate (see for instance \cite{Ruiz:2016rai}).

\subsection{Diagnostics}
\label{sec:diagnostics}
Throughout the entire numerical evolution of our configurations, we monitor various quantities to 
ensure the reliability of our simulations and to track the evolution of the BNS remnants. A brief description of these quantities is provided below.
\paragraph{\bf Constraints:} Throughout the evolution, we monitor the $L_2$ norm of the normalized Hamiltonian and the normalized momentum constraints as defined in Eqs.~(40)-(43) of~\cite{Etienne:2007jg}. In all simulations the Hamiltonian constraint violations remain smaller than $\sim0.05\%$ during the inspiral. In the case of BH formation, the constraints peak at $\sim1.6 \%$ and then relax to $\sim0.1\%$ as the system reaches a steady-state. The momentum constraint violations remain smaller than $\sim1\%$ during  the inspiral, peak at $\sim10.5\%$ during BH formation (if applicable) and then relax to $\sim0.8\%$ as the system reaches a steady-state. These values are similar to those previously reported in our long-term, hydrodynamic simulations of spinning BNSs modeled using the  SLy and ALF2 EoSs~\cite{Tsokaros:2019anx}.
\paragraph{\bf BH Horizons:} 
In cases where the binary undergoes delayed collapse, we employ the {\tt AHFinderDirect} thorn~\cite{Thornburg:2003sf} to 
track the apparent  horizon of the BH. Additionally, we estimate the BH mass and its dimensionless spin using the isolated 
horizon formalism  as described in~\cite{Dreyer:2002mx}. 
\paragraph{\bf Gravitational waves:} 
To compute the Weyl scalar $\Psi_4$, we use the {\tt Psikadelia} 
thorn, decomposing it into $s=-2$  spin-weighted spherical harmonic modes. Using Eqs. (2.8), (2.11), and (2.13) in~\cite{Ruiz:2007yx},
we calculate the total flux of energy and angular momentum carried away by GWs at ten extraction radii, ranging from 
$r_{\rm min}\approx120M$ and  $r_{\rm max}\approx840M$ and perform an extrapolation to infinity.
We find that between~$\sim 1.5\%$ and $\sim 3.0\%$ of the total energy of our binaries is radiated away 
in the form of gravitational radiation, while between $\sim 20.0\%$ and $\sim 27.9\%$ of the angular momentum is 
radiated (see~Table~\ref{tab:results}).

Following~\cite{Maione:2017aux,DePietri:2019mti}, we analyze the post-merger gravitational wave signals
using a modified version of the Prony's method, which allows us to accurately characterize the main gravitational wave mode and their frequencies. We use the {\tt ESPRIT} (Estimation of Signal Parameters via Rotational Invariance Techniques) variant of Prony's method~\cite{https://doi.org/10.1002/gamm.201410011, POTTS20131024}, which is robust against noise and involves fitting a larger number of complex exponentials to the data. Our methodology uses a moving window interval of 
$4\, {\rm ms}$ to perform the analysis over time. This process includes performing Fast Fourier Transforms (FFT) on segments of the signal, extracting Fourier amplitudes at selected fixed frequencies, and correlating these with specific oscillation modes. We report, in Table~\ref{tab:results}, the fundamental frequency associated with the dominant $l=m=2$ oscillation mode $f_2$, its transient value $f_{2i}$, typically computed during the first few milliseconds after merger, along with the $f_{\rm max}$, which corresponds to the instantaneous GW frequency at merger, i.e. at the GW peak amplitude (see the vertical dashed lines in Fig.~\ref{fig:waveforms}) \cite{Takami:2014tva}.
Furthermore, we obtain the time-frequency spectrograms of the $l=m=2$ 
component of the GW strain for all models, and plot them against the active modes that are responsible for the most part of the GW emission. We also performed a density distribution FFT and extracted the Fourier amplitudes at specific fixed frequencies corresponding to the main modes observed in the spectrograms. These amplitudes represent the eigenfunctions of the oscillation modes~\cite{DePietri:2019mti}.

We use time-domain waveforms from the binaries in Table~\ref{table: initial data} that do not collapse to a BH, injecting them into Gaussian noise 
consistent  with the sensitivity of the third-generation Einstein Telescope (ET). The signals are then reconstructed using {\tt BayesWave}~\cite{Cornish:2015},  a Bayesian data analysis algorithm that employs a model-agnostic approach for signal recovery.
\paragraph{\bf Global Conservation:}
We check the conservation of the ADM mass and the ADM angular momentum along with the conservation of the rest-mass 
given by the expressions (21)-(23) in~\cite{Etienne:2011ea}, along with the conservation of the rest-mass computed through 
Eq.~(23) in~\cite{Etienne:2011ea}.
The ADM mass is conserved to $< 1\%$,  the angular momentum to within $\lesssim 3\%$, and the rest-mass
within~$\lesssim 1.2\%$.
\begin{figure*}
    \includegraphics[scale=0.425]{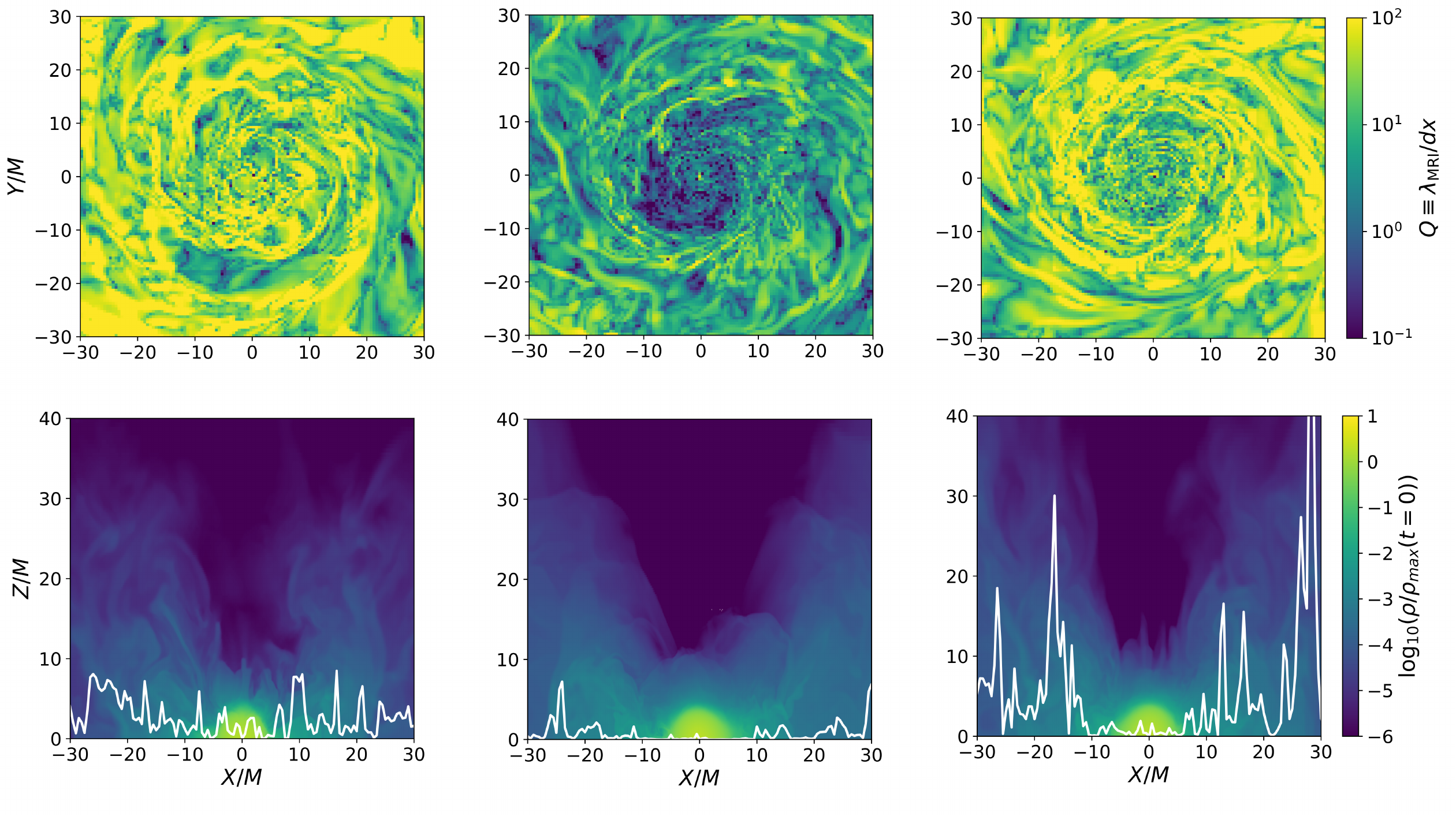}
     \caption{Top panel: Pseudocolor plot of the quality factor $Q_{\rm MRI}=\lambda_{\rm MRI}/dx$ on the equatorial (xy) plane
    for the remnant of the SLy BNS at~$\sim 3\,{\rm ms}$ after merger with: i) a pure initial poloidal field (left); a pulsar-like + mixed magnetic field (middle); and an interior mixed field transitioning smoothly to an external pulsar-like magnetic field at $0.95 R_{\rm NS}$ (right). Bottom panel: Rest-mass density normalized to its initial maximum value (logarithmic scale) on a meridional (xz) slice, along with the $\lambda_{\rm MRI}$ (white line) for the same cases as in the top panels. For the most part the $\lambda_{\rm MRI}$ fits within the remnant.}
    \label{fig:mri_c3}
\end{figure*}

\paragraph{\bf Electromagnetic signatures:}
We compute the magnetic energy \cite{Etienne:2011ea}
\begin{equation} 
\label{eq:mag-energy}
    \mathcal{M}=\int u^\mu u^\nu T_{\mu\nu}^{\rm(EM)}\,dV\,,
\end{equation} and monitor it throughout the simulations.
where $u^\mu$ is the  fluid 4-velocity and $T_{\mu\nu}^{\rm(EM)}=b^2u_\mu u_\nu+b^2 g_{\mu\nu}/2-b_\mu b_\nu$ is the 
stress-energy tensor associated with the magnetic field, and $dV$ is proper volume element on a $t =\rm constant$ spatial 
slice. Here $b^2=b^\mu b_\mu$ is the magnetic field measured in the fluid’s comoving frame~\cite{Etienne:2011ea}.

We evaluate the outflow of matter
\begin{align} \label{eq:ejecta-mass}
    M_{\rm esc} = &\int_{r>r_0}\rho_*\Theta(-u_t-1)\Theta(v^r){\rm d}x^3 \\ \nonumber
    &+\int_{t'=0}^t\int_{\delta\mathcal{D}}\rho_*\Theta(-u_t-1)\Theta(v^r)v^i{\rm d}S_i{\rm d}t' \, ,
\end{align}
where $\rho_*=\sqrt{-g}\rho_0u^t$, with $g$ the determinant of the 4-metric. The surface element on the sphere 
is denoted by ${\rm d}S_i$. Heaviside functions $\Theta$ are used to include only the material with positive 
specific energy $E=-u_t-1$ and positive radial velocity $v^r$. The second term is the contribution from the 
rest mass leaving the boundary of the simulation domain $\delta\mathcal{D}$.  We measured the outflow of matter 
at different radii between $r_0=30M$ and $100M$ and find that the difference between them is less than 
$\lesssim 3.6\%$ consistent with the values found in BNS simulations modeled with a SLy EoS in~\cite{Bamber:2024kfb}. 

In addition, we compute the fluid luminosity, defined as
\begin{equation}\label{eq:fluid-luminosity}
    L_{\rm fluid} = \int \sqrt{-g}(-T_t^{i({\rm fluid})}-\rho_0 u^i)dS_i \, ,
\end{equation}
where $T_t^{i({\rm fluid})}$ is the projection of the stress-energy tensor of a perfect fluid, and the EM Poynting 
luminosity as
\begin{equation}\label{eq:poynting-luminosity}
    L_{\rm EM} = -\int \sqrt{-g} T_t^{i({\rm EM})} dS_i \, ,
\end{equation}
over ten spherical surfaces with radii ranging from $120$ to $840M$. Furthermore, for the SLy $M=2.70M_{\odot}$ cases which form black holes, we compute the 
expected electromagnetic luminosity for a jet powered by the steady-state BZ mechanism, which can 
be estimated as \cite{Thorne:1986iy} 

\begin{equation}\label{eq:BZ-lum}
    L_{\rm BZ} \sim 10^{52}\left(\frac{\chi_{\rm\tiny BH}}{0.64}\right)^2\left(\frac{M_{\rm BH}}{2.5\Msol}
    \right)^2\left(\frac{B_{\rm pol}}{10^{16}\,\rm G}\right)^2\, \rm erg\,s^{-1} \, ,
\end{equation}
where $\chi_{\tiny\rm {BH}}$ and $M_{\rm BH}$ are the dimensionless spin and mass of the BH remnant, respectively, and $B_{\rm pol}$ is the strength of the magnetic field at the BH pole.

We compute the effective Shakura–Sunyaev parameter associated with the effective viscosity due to magnetic stresses
\begin{equation}
    \alpha_{\rm SS} \sim \frac{\rm magnetic\,stress}{\rm pressure} =
    \frac{T^{\rm EM}_{\hat{r}\hat{\phi}}}{P} \,,
\end{equation}
where $T^{\rm EM}_{\hat{r}\hat{\phi}}={e}_{\hat{r}}^\mu {e}_{\hat{\phi}}^\nu T_{\mu\nu}^{\rm EM}$ is the component of the
electromagnetic stress-energy tensor in the local comoving frame and $e_{\hat{i}}^\mu$ the corresponding basis of local tetrads
(see Eq.~(26) in \cite{2010MNRAS.408..752P}). We find that $\alpha_{\rm SS}\sim 10^{-3}$ inside the remnant, and $\sim10^{-2}$ in the disk. However, note that 
it has been shown this value depends on resolution \cite{2011ApJ...738...84H,Hawley:2013lga,Kiuchi:2017zzg}. 
\paragraph{\bf Remnant properties:} 
We estimate the coordinate angular velocity of the metastable HMNS as
\begin{equation}\label{eq:ang-vel}
    \Omega \coloneqq \frac{u^\phi}{u^t} \, ,
\end{equation}
with $u^\phi$ and $u^t$ the azimuthal and time components of the four-velocity.

Following \cite{Paschalidis:2011ez}, we compute the temperature $T$  of the HMNS via
\begin{equation}\label{eq:temp}
    \epsilon_{\rm th} =\frac{3k_BT}{2m_{\rm n}}+f\frac{aT^4}{\rho_0} \, ,
\end{equation}
where $\epsilon_{\rm th}=\epsilon-\epsilon_{\rm cold}$ is the specific thermal energy density, while  $\epsilon$ and $\epsilon_{\rm cold}$ are the specific energy density and its cold part. Here, $k_B$ is Boltzmann's constant, $m_{\rm n}$ is the mass of the neutron, and $a$ is the radiation constant. The first term approximates the thermal energy of the nucleons, while the second term accounts for the thermal energy contributed by relativistic particles. The factor $f$ reflects the number of species of relativistic particles that contribute to the thermal energy, and we assume $f=11/4$, i.e. we account only for the contribution of electrons and positrons to the thermal radiation. At temperatures exceeding tens of ${\rm MeV}$, thermal neutrinos are produced abundantly \cite{Cusinato:2021zin}, such that the assumption of $f=11/4$ is no longer valid. A more consistent treatment would instead rely on tabulated equations of state that incorporate thermal, compositional, and neutrino contributions. However, the simple model using a single value of $f$ suffices as a first approximation.

\paragraph{\bf The Brunt-V\"ais\"ala frequency:}
{To probe the convective stability of our BNS remnants and understand their internal dynamics, we estimate the Brunt-V\"ais\"ala frequency~$\mathcal{N}$, which quantifies the oscillation frequency of a vertically displaced fluid element around its equilibrium position \cite{Cox80, Tassoul_2000}, via
\begin{equation} \label{eq:brunt-approx}
    \mathcal{N}^2\propto \mathcal{B}_i\mathcal{G}^i \, ,
\end{equation}
where $\mathcal{B}_i$ is the Ledoux discriminant, and $\mathcal{G}^i$ is the gravitational acceleration. Convective stability has previously been studied for non-magnetized BNS remnants \cite{DePietri:2019mti,Gao:2025nfj,Guerra2025prep}. It has also been studied in the linear regime of the MRI (see, e.g. \cite{Cerda-Duran:2007fvm}). In the nonlinear regime of the MRI, numerical simulations of shearing boxes have been performed in three dimensions in the context of accretion disks \cite{1995ApJ...440..742H}. Here, we use the Brunt-V\"ais\"ala frequency as a proxy to study the non-linear regime in the context of magnetized BNS remnants for the first time. A more detailed description of our approach and the validity of our approximations is provided in Appendix \ref{appendix:convective-stability}.}

\section{Results}
\label{sec:results}

\begin{table*}[]
\centering
\caption{Summary of the key properties of the binary mergers. Here, the first column indicates whether the MRI is fully resolved, $t_{\rm merg}$ is the merger time in ${\rm ms}$, 
$f_{2i}$ is the instantaneous frequency at the time of the merger associated with the dominant oscillation mode 
($l=m=2$), and $f_2$ is the frequency once the binary remnant has settled into a quasi-stable state. All values of frequency are reported in ${\rm kHz}$.
The $f_2$ values are not reported for cases that collapse $\sim4\,{\rm ms}$ after merger, as the frequency increases 
as the remnant becomes more compact, making it impossible to define a single dominant frequency.
The fractions of the total energy $M$ and angular momentum $J$ carried off by GWs are $E_{\rm GW}/M$ and $J_{\rm GW}/J$, respectively.
$M_{\rm esc}$ is the escaping rest-mass (ejecta) calculated using Eq.~\eqref{eq:ejecta-mass} at $20\,{\rm ms}$ 
after merger. $L_{\rm EM}$ is the Poynting luminosity in ${\rm erg\,s}^{-1}$ $20\,{\rm ms}$ after merger. $L_{\rm knova}$, 
$\tau_{\rm peak}$, and $T_{\rm peak}$ denote, respectively, the estimated peak EM luminosity in $\rm erg/s$, 
rise time in days, and temperature of the potential kilonova arising from the sub-relativistic ejecta in ${\rm K}$ 
(see Eqs.~\eqref{eq:kilonova_L},~\eqref{eq:kilonova_t}, and ~\eqref{eq:kilonova_T}). These
are calculated from the ejecta mass and the averaged ejecta velocity $v_{\rm eje}$ of the outflow computed at 
$\sim 160\,{\rm km}$ from the remnant. $t_{\rm BH}$ is the black hole formation time in ${\rm ms}$ since merger. 
$M_{\rm BH}$ and $\chi_{\rm BH}$ are the mass and spin of the black holes at the end of the simulation, calculated using 
the isolated horizon formalism. A dash symbol indicates ``not applicable". Finally, the last column shows the final time post-merger of the simulations.
}
\resizebox{\linewidth}{!}{%
\begin{tabular}{c|ccccccccccccccccccc}
\hline\hline
Label  & $Q\gtrsim10$ & $t_{\rm merg}$ & $f_{2i}$ & $f_2$ & $f_{\rm max}$ & $E_{\rm GW}/M$ & $J_{\rm GW}/J$ & $M_{\rm esc}$ & $\langle v_{\rm eje}\rangle$ & $L_{\rm EM}$ & $L_{\rm knova}$ & $\tau_{\rm peak}$ & $T_{\rm peak}$ & Fate & $t_{\rm BH}$     & $M_{\rm BH}$ & $\chi_{\rm BH}$ & $t_{\rm final}$ \\ \hline
SLy\_H    & - & 7.04          &  3.61     & 3.76       &    1.97 & 2.69\% & 27.9\% & 0.49\%         & 0.21                         & -            & $10^{40.63}$    & 3.65              & $10^{3.28}$   & BH & 15.00    & 2.49 & 0.67 & 29.4\\
SLy\_P  & \cmark & 7.04          & -         & -          &  1.98     & 1.83\% & 21.3\% & 1.13\%         & 0.30                         & $10^{51.08}$ & $10^{40.71}$    & 4.64              & $10^{3.28}$   & BH & 3.34     & 2.59 &  0.74 & 21.6\\
SLy\_SP  & \xmark & 7.04          &   4.46    & 3.82       &   1.96    & 2.59\% & 27.2\% & 0.47\%         & 0.32                         & $10^{50.29}$ & $10^{40.59}$    & 2.90              & $10^{3.34}$   & BH & 15.41    & 2.48 & 0.66   & 20.6\\
SLy\_T$_{0.95}$ & \cmark & 7.02          & -         & -          &    1.96 & 2.07\% & 22.9\% & 0.83\%         & 0.25                         & $10^{50.87}$ & $10^{40.61}$    & 4.35              & $10^{3.31}$    & BH & 3.70    & 2.59 &  0.74   & 21.3\\
SLy\_T$_{0.5}$  & \cmark & 7.04          &  3.84     & 3.93       &   1.97    & 2.41\% & 25.8\% & 1.15\%         & 0.20                         & $10^{50.92}$ & $10^{40.59}$    & 5.77              & $10^{3.30}$   & BH & 13.17    & 2.46 &  0.64   & 20.9\\ \hline
WFF1\_H  & - & 3.26          &  3.88     & 4.00       &   2.32    & 3.01\% & 27.9\% & 0.53\%         & 0.21                         & -            & $10^{40.49}$    & 3.74              & $10^{3.35}$   & HMNS & -    & -& - &  40.2\\
WFF1\_P & \cmark & 3.25          &  4.04     & 4.11       &   2.33    & 2.49\%  & 24.0\% &3.50\%         & 0.22                         & $10^{51.84}$ & $10^{40.79}$    & 9.34              & $10^{3.21}$   & HMNS & -     & -&  - & 40.3\\
WFF1\_SP  &  \xmark & 3.26          &  3.82     & 3.83       &   2.34    & 2.83\% & 26.8\% &0.52\%         & 0.20                         & $10^{50.84}$ & $10^{40.62}$    & 3.79              & $10^{3.28}$    & HMNS & -     & -& -  & 40.1\\
WFF1\_T$_{0.95}$  & \cmark & 3.25          &  3.72     &  3.83      &    2.34  & 2.47\% & 24.2\% & 1.35\%         & 0.20                         & $10^{51.60}$ & $10^{40.62}$    & 6.15              & $10^{3.29}$    & HMNS & -     & -& - & 40.0\\
WFF1\_T$_{0.5}$ & \cmark  & 3.26          &  3.92     &  4.04      &    2.33  & 2.73\% & 26.0\% & 1.55\%         & 0.21                         & $10^{51.55}$ & $10^{40.66}$    & 6.35              & $10^{3.27}$    & HMNS & -   & -& -   & 40.8\\ \hline
SLy2.54\_P  &  \cmark &   6.77      &    3.34   &   3.41     &   1.92    & 1.55\% & 20.0\% &2.57\%        &    0.22                     & $10^{52.80}$ &   $10^{40.89}$   &   7.92            &   $10^{3.17}$  & SMNS & -    & -& - & 50.6\\ \hline\hline
\end{tabular}
}
\label{tab:results}
\end{table*}
The dynamics of the inspiral phase of the BNS cases in Table~\ref{table: initial magnetic field} is roughly the same. 
The stars begin in a quasi-circular orbit and gradually lose energy and angular momentum through GW 
radiation, causing their coordinate separation to shrink. As they approach each other, tidal forces 
become significant, distorting the stars until they merge, forming a dual-core transient configuration. 
These cores orbit each other within a hot, nuclear-density envelope and eventually merge, forming 
a highly differentially rotating central core surrounded by a low-density cloud of matter. Depending on 
the EoS, the total mass of the system, and the magnetic field configuration, the binary remnant undergoes 
collapse into a BH or forms a metastable HMNS remnant with a lifetime longer than the duration of our 
simulations. Consistent with~\cite{Bamber:2024qzi,Ciolfi:2020hgg}, we verify that, depending on the 
initial poloidal magnetic field strength  in the NSs, the BH + disk remnant launches a magnetically-driven 
outflow compatible with sGRBs.  Further studies of the baryon pollution in the funnel are required to understand 
if there are other mechanisms for efficient particle acceleration in the stable NS remnants scenario~\cite{Kiuchi:2023obe}.
Key physical properties characterizing the evolution of the binaries are summarized in Table~\ref{tab:results}.

\subsection{Post-merger evolution}
\label{subsec:post-merger_evol}

\subsubsection{Undergoing delayed collapse to a BH} 
Consistent with previous GRMHD studies using similar magnetic field strengths and resolutions~\cite{Bamber:2024kfb,Ruiz:2021qmm},
we find that SLy binaries with an ADM mass of $2.7M_\odot$ merge to form a highly differentially rotating remnant with a total
rest mass of $\sim 3M_\odot$.  Large-scale magnetic fields efficiently redistribute angular momentum through magnetic tension \cite{Shapiro:2000zh}, which eventually causes the formation of a uniformly rotating core surrounded by a Keplerian
cloud of low density material. In addition, MRI-active layers outside the core may still contribute indirectly through non-local angular momentum transport, coupling different regions of the remnant. Other MHD instabilities (e.g. Tayler-type instabilities) may also play a role in enhancing magnetic stresses. Since the core exceeds the supramassive threshold for the SLy EoS ($2.96M_\odot$)
\cite{Bamber:2024kfb,Kolsch:2021lub}, the remnant ultimately collapses to a spinning black hole wrapped by an accretion disk.

\begin{figure}
    \centering
    \includegraphics[width=\linewidth]{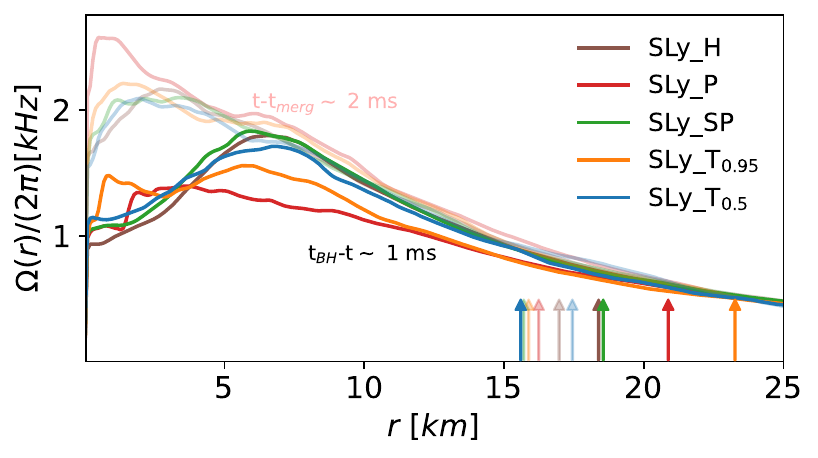}
    \caption{Average azimuthal angular velocity profiles of the binary remnant that undergo collapse (see Table~\ref{tab:results})
     at  $\sim 2\, {\rm ms}$ following merger (faded lines) and at $1\,{\rm ms}$ before BH formation (bold lines), which corresponds to $t-t_{\rm merg} > 13 \,{\rm ms}$ for configurations {\bf{H}}, {\bf{SP}}, and {\bf{T}}$_{0.5}$, and $t-t_{\rm merg} \sim 3 \,{\rm ms}$ for {\bf{P}} and {\bf{T}}$_{0.95}$. Arrows mark the rest
     mass density contour at $\rho_0=10^{-3}
    \rho_{0,\rm max}(t=0)$, which gives a rough measure of the size of the remnant.}
    \label{fig:SLy-rot-profile}
\end{figure}

Depending on the initial magnetic field configuration, the lifetime of the remnant ranges from
$\sim 4$ to $\sim 15\,{\rm ms}$ (see Tables~\ref{table: initial magnetic field} and~\ref{tab:results}).  In particular, the {\bf{SP}} case for which the MRI is not reliably resolved, has a lifetime
comparable to the non-magnetized scenario. Fig.~\ref{fig:mri_c3} shows the MRI quality factor  $Q_{\rm MRI} = \lambda_{\rm MRI}/dx$ 
on the equatorial plane at $t-t_{\rm merg} \sim 3\,{\rm ms}$ for the {\bf{P}}, {\bf{SP}}, and $\mathbf{T}_{0.95}$ configurations. 
Whether the MRI is fully resolved is also explicitly given in Table~\ref{tab:results}. Figure~\ref{fig:SLy-rot-profile} shows the angular velocity profiles of the merger remnants at 2 ms after merger and 1 ms before collapse. At these early times, the remnant is still in a strongly dynamical and oscillatory phase. Nevertheless, for the longer-lived cases ({\bf{H}}, {\bf{SP}}, {\bf{T}}$_{0.5}$) the profiles indicate the emergence, before collapse, of a slowly varying rotational core surrounded by a low-density envelope with approximately Keplerian rotation. While magnetic angular momentum redistribution in the dense core is expected to occur on Alfvén timescales longer than a few milliseconds, such redistribution can be effective in the outer, low-density regions, contributing to the observed evolution of the rotation profile.

\begin{figure}
    \centering
    \includegraphics{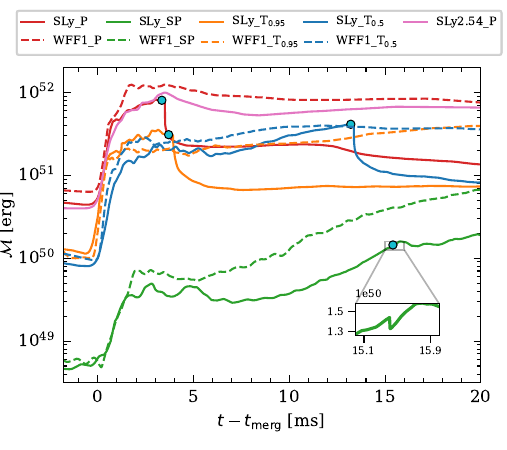}
    \caption{Evolution of the magnetic energy $\mathcal{M}$ as computed via Eq.~\eqref{eq:mag-energy} for cases listed in Table~\ref{table: initial data} endowed with
    the magnetic field configurations in Sec.~\ref{sec:magnetic-fields}. The cyan circles mark the BH formation time.
    The inset highlights a slight decrease in $\mathcal{M}$ just after BH formation in the SLy\_SP case. However, its magnetic energy continues to increase beyond this point.}
    \label{fig:magnetic-energy}
\end{figure}

The total magnetic energy, including contributions from EM outflows leaving the numerical domain, is displayed in Fig.~\ref{fig:magnetic-energy}. Although the initial poloidal and toroidal magnetic energies are reported (see Fig.~\ref{table: initial magnetic field}), their individual evolution was not analyzed in the present work. Such an analysis is left for future studies. Nevertheless, the overall amplification of the magnetic energy, together with the expected winding of the magnetic field, suggests a progressive growth of the toroidal component during the evolution.
Following  merger, the magnetic energy increases by more than an order of magnitude within the first $\sim 1.5\,{\rm ms}$, consistent with amplification
driven by the KHI at the shear interface~\cite{Bamber:2024kfb}. The total magnetic energy increases exponentially, $\propto \exp(2\gamma_{\rm KH}(t-t_{\rm merg}))$, with $\gamma_{\rm KH}$ the characteristic growth rate. Consistent with~\cite{Bamber:2024kfb}, we observe that the relative amplification 
growth rate is $\gamma_{\rm KH} \sim 1.38\,{\rm ms}^{-1}$.	As pointed out by~\cite{Bamber:2024kfb}, this magnetic energy growth rate $2\gamma_{\rm KH} 
\sim 2\,{\rm ms}^{-1}$ is several orders of magnitude lower than the growth rate predicted by linear perturbation theory $\sigma_{\rm KH} 
\sim 10^2\,{\rm ms}^{-1}$  based on $\Delta v \sim 0.1c$ and shear layer width $d \sim 400\,{\rm m}$.  
This is likely due to lack of resolution, which partially resolves the fastest-growing 
KH modes, and/or the relatively low efficiency of the instability in converting kinetic shear energy into magnetic energy.
Studies of spinning and irrotational BNSs using numerical resolutions of a few meters are needed to estimate the growth rate in the non-linear regime.

As the remnant begins to settle, high velocity gradients, responsible for driving the KH instability, diminish and the MRI and magnetic winding
take over causing a change in the growth rate of the magnetic field. Compression of frozen-in magnetic field in the remnant as it becomes more compact will also increase the total energy. However, the change in compactness before collapse is relatively small, so the compression induced change is likely subdominant compared to the other mechanisms. The magnetic energy in the {\bf{P}} 
and $\mathbf{T}_{0.95}$ remnants peaks roughly at $\gtrsim 10^{51}\,{\rm erg}$ just before BH formation followed by a sharp drop as highly magnetized 
material in the core is accreted into the BH. Afterward, the magnetic energy decreases more gradually, continuing until the end of the simulation 
due to the ongoing infall of surrounding material. 
In the {\bf{SP}} and {\bf{T}}$_{0.5}$ cases, our least magnetized models, we observe that after $t-t_{\rm merg}\sim 4\,{\rm ms}$ when the 
KHI weakens, the magnetic field initially grows exponentially mainly due to the MRI and magnetic winding,  then it transitions to a non-linear 
regime before eventually saturating. Following this saturation, the magnetic energy declines, which is particularly evident in {\bf{SP}} after 
$t-t_{\rm merg}\sim 5\,{\rm ms}$. This decline, also reported in higher-resolution simulations (see 
e.g.~\cite{Aguilera-Miret:2023qih,Kiuchi:2023obe}),  has been attributed to the conversion of magnetic energy into kinetic energy 
during the acceleration of ejecta or to the collapse of an unstable magnetic configuration~\cite{Cheong:2024stz}. We note that the angular velocity profile
at $\sim 1\,{\rm ms}$ before BH formation is higher in these two cases compared to the more strongly magnetized ones (see Fig.~\ref{fig:magnetic-energy}).
By $t-t_{\rm merg}\sim 6\,{\rm ms}$ the total magnetic energy gradually begins to rise again due to the influence of magnetic instabilities, although this 
increase is partially balanced by the conversion of magnetic energy into the kinetic energy of the accelerated gas. Following BH formation, the magnetic 
energy in {\bf{T}}$_{0.5}$ drops sharply as its highly magnetized central core is swallowed by the newly formed BH. Afterward, the magnetic energy slowly 
decreases as infall material is accreted. In contrast, during BH formation the magnetic energy in {\bf{SP}} exhibits a slight decrease (see inset in Fig. 
\ref{fig:magnetic-energy}), followed by a continuous increase. Although the value of the initial magnetic-to-gas pressure ratio within the NSs is constant across all cases, {\bf{SP}} has the lowest initial magnetization energy and hence the wavelength of the fastest growing MRI mode in the  
HMNS epoch is relatively short. Following BH formation, the magnetic energy in this case continues to grow until the termination of our simulation, 
driven by both the MRI and magnetic winding. Since the MRI characteristic wavelength scales with $\rho_0^{-1/2}$ (see Eq.~(\ref{eq:MRI_grid})), 
it is resolved after BH formation with the adopted resolution.

As shown in Table~\ref{tab:results}, the mass $M_{\rm BH}$ of the BH remnant and its dimensionless spin $\chi_{\rm BH}$ 
is roughly the same across different cases, with values of $M_{\rm BH} \sim 2.5$ and $\chi_{\rm BH} \sim 0.7$ near the 
termination of the simulations. This is expected as the physical properties of the BH remnant are mainly 
determined by the the mass of the system and the orbital angular momentum of the system at merger.

\begin{figure}
    \centering
    \includegraphics[width=1.0\linewidth]{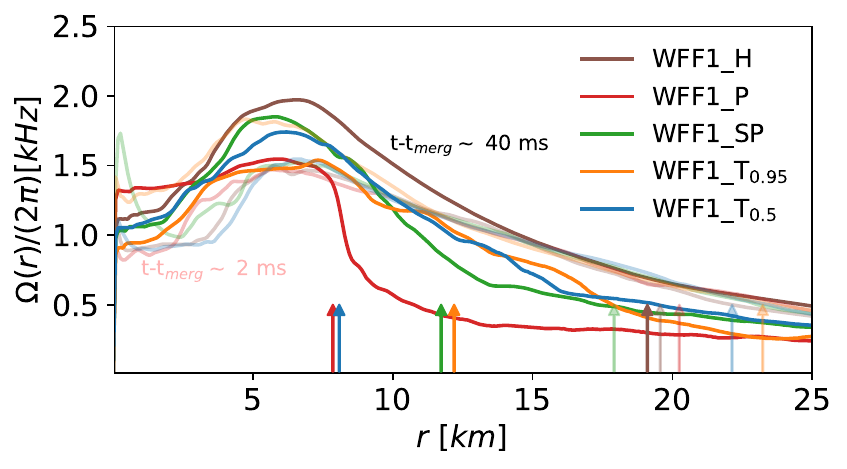}
    \caption{Average azimuthal angular velocity profiles of the stable binary remnants (see Table~\ref{tab:results}) in
    at  $\sim 2\,{\rm ms}$ following merger (faded lines) and near to the termination of our simulations (bold lines). 
    Arrows marks the rest mass density contour at $\rho_0=10^{-3}
    \rho_{0,\rm max}(t=0)$, which gives a rough measure of the size of the remnant.}
    \label{fig:rot-WFF1}
\end{figure}

\begin{figure*}
\centering
\includegraphics[scale=0.3]{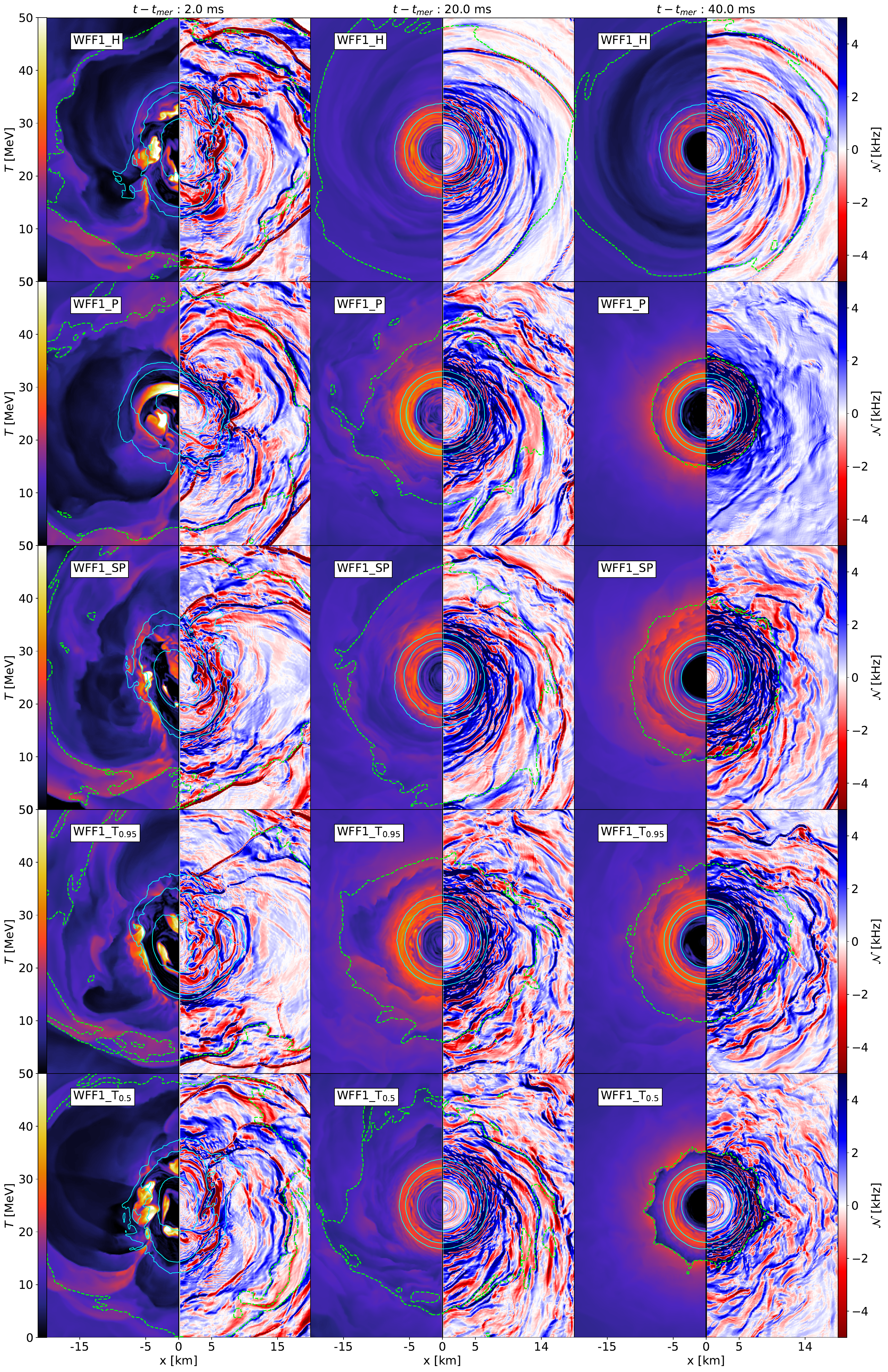}
\caption{Snapshots of the temperature in MeV calculated via Eq.~\eqref{eq:temp} (left panels) and the Brunt-V\"ais\"ala frequency in kHz
calculated via Eqs.~\eqref{eq:Bi}~and \eqref{eq:Gi}  (right panels) for the WFF1 BNS remnants with varying magnetic field content on the 
equatorial plane at selected times. Cyan solid lines show different rest-mass density contours corresponding to the boundaries between segments 
of the piecewise equation of state used to model the star, corresponding to densities $1.76\times10^{14} \,{\rm g/cm}^3$, $5.01\times10^{14} \,{\rm g/cm}^3$, and $10^{15} \,{\rm g/cm}^3$. The green dashed line marks the bulk of the star, defined as $10^{-3}\rho_0^{\rm max}$. }
\label{fig:tempvsbrunt}
\end{figure*}

\begin{figure*}
\centering
\includegraphics[scale=0.5]{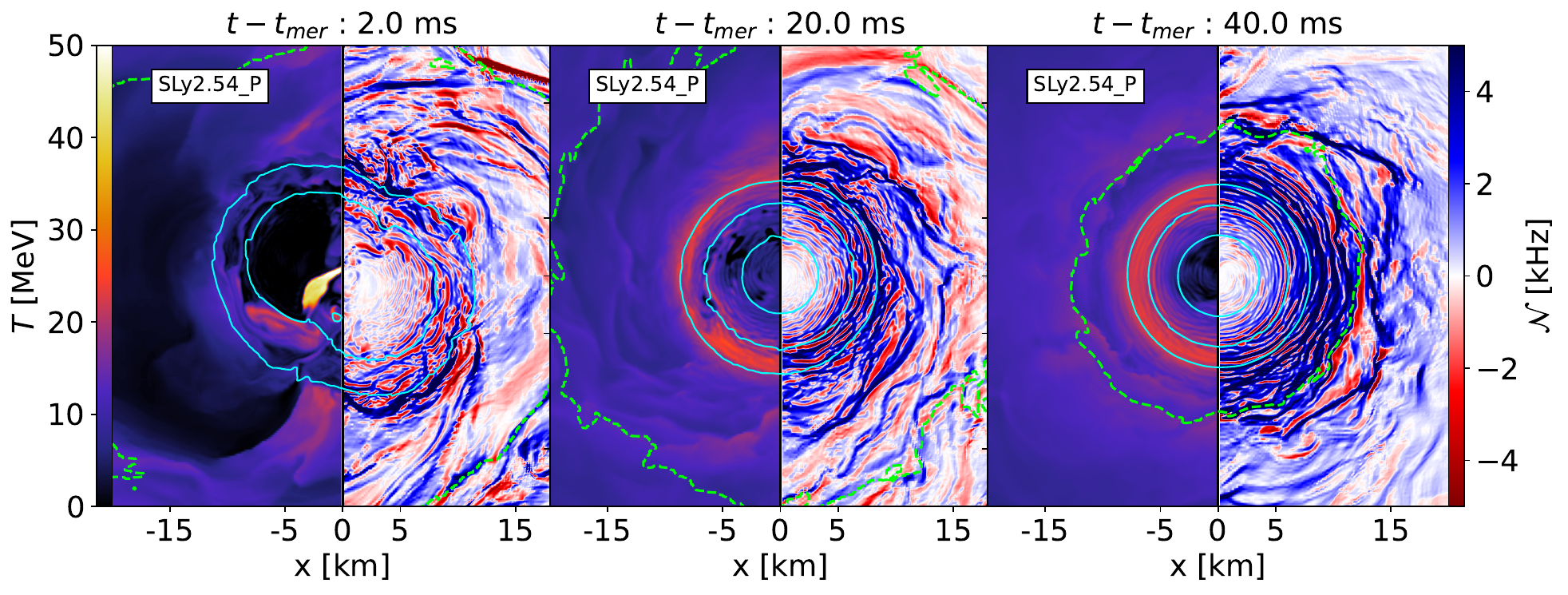}
\caption{Same as in Fig.~\ref{fig:tempvsbrunt}, but for the remnant of the SLy2.54\_P case. Here, cyan lines correspond to densities $1.46\times10^{14}\,{\rm g/cm}^3$, $5.01\times10^{14}\,{\rm g/cm}^3$, $10^{15}\,{\rm g/cm}^3$.}
\label{fig:tempvsbruntSLy}
\end{figure*}

\subsubsection{(Meta)stable remnants} 
For the SLy2.54 and WFF1 binaries, the merger remnants are a stable SMNS~\cite{Bamber:2024kfb} and  long-lived HMNS remnants, 
respectively, with lifetime exceeding our simulated time $t-t_{\rm merg}>40\,{\rm ms}$.  The early evolution of these systems 
closely resembles that discussed in the previous section, with the key distinction being the nature of the remnant formed in 
each case. Following merger, the magnetic energy increases by more than an order of magnitude within the first $\sim 1.5\,{\rm ms}$. 
After that point, the magnetic energy remains roughly constant in all cases but WFF1$\_$SP, for which we observe that the magnetic energy continues to grow until the termination of our simulation, likely driven by both the MRI and magnetic winding. 

Fig.~\ref{fig:rot-WFF1} displays the rotational profiles of the WFF1 remnants as computed via Eq.~\eqref{eq:ang-vel} shortly after merger ($t-t_{\rm merg}\sim 2\,{\rm ms}$) and near the termination of our simulations ($t-t_{\rm merg}\sim 40\,{\rm ms}$). 
Notice how, in particular for {\bf{P}}, the curve flattens as time goes on. This flattening of the rotational profile is due to the 
transport of angular momentum induced by the magnetic viscosity. Naturally, as the employed resolution for the {\bf{SP}} case is not
enough to fully resolve the MRI, we observe that this case retains a significant degree of differential rotation even at late times. By the same logic, the non-magnetized case is, as expected, the one retaining the largest degree of differential rotation. The pulsar-like configuration is almost rigidly rotating (see solid red line in Fig.~\ref{fig:rot-WFF1}, where the solid arrow indicates the radius of the corresponding remnant). We observe that the efficiency of the angular momentum redistribution correlates to the maximum initial magnetic field strength (see e.g. Table~\ref{table: initial magnetic field}), which is expected as the fastest-growing wavelength of the MRI is proportional to the magnetic field strength (see Eq.~\eqref{eq:MRI_grid}).

Figs.~\ref{fig:tempvsbrunt} and~\ref{fig:tempvsbruntSLy} show the temperature of the remnants calculated with Eq.~\eqref{eq:temp}, on the left of each panel, and the Brunt-V\"ais\"al\"a frequency $\mathcal{N}$ given by Eq.~\eqref{eq:brunt-approx} (see Appendix~\ref{appendix:convective-stability} for more details), on the right. Cyan lines display different rest-mass density contours corresponding to the boundaries between segments of the piecewise equation of state used to model the star, while the green dashed line marks
the surface of the star, which we define as $10^{-3}\rho_0^{\rm max}$.

Immediately after merger, the remnants reach maximum temperatures of $\gtrsim 50\,{\rm MeV}$, due to the strong shock-heating. By $20-40\,{\rm ms}$ after merger, the cores cool and contract, while the outer layers retain heat in a ring structure, consistent with findings from previous studies of non-magnetized BNS mergers (see, e.g. ~\cite{Kastaun:2016yaf}). Here, we find that the properties of this ring are not only affected by the presence of a magnetic field, but also by its topology.

The hot ring can be attributed to the differential rotation profile present in the remnant shortly after merger. During this epoch, the outer layers possess higher specific angular momentum than the inner core 
(see  Fig.~\ref{fig:rot-WFF1}). As a result, angular momentum transport and centrifugal support push material outward, leading to the formation
of a ``centrifugal barrier", a region where the outward centrifugal force balances gravitational attraction.
As the merger proceeds, shock waves are generated as a result of the collision and compression of matter. These shocks heat the material they pass through, 
and as they propagate outward, they reach the regions where matter is accumulating due to the centrifugal barrier. The material in these regions 
reaches temperatures of $\sim30-40\,\rm MeV$. Note that in the calculation of the temperature, we do not consider the presence of thermal neutrinos. If we were to consider their presence, the temperature would be lower. Furthermore, our simulations do not include a treatment for neutrino transport, and previous simulations with such treatments have shown that the temperature decreases when introducing neutrinos \cite{Zappa:2022rpd}. Therefore, we would expect the true temperature of BNS remnants to be a few percent lower than what we report here.

In magnetized cases, magnetic instabilities amplify the magnetic field by extracting energy from differential rotation. Part of this magnetic energy is subsequently converted into heat through dissipation processes, sustaining high temperatures for longer. In contrast, the non-magnetized remnant lacks these magnetic effects and, as a result, it cools down and is the coolest configuration at $\sim 40\,{\rm ms}$ after the merger (see the right column in Fig.~\ref{fig:tempvsbrunt}).

At early times post-merger, the Brunt-V\"ais\"al\"a frequency shows a complex pattern of stable and unstable regions, driven by violent oscillations and inhomogeneous heating. At $20\,{\rm ms}$ after merger, the structure becomes more coherent, with extended stable regions. At $40\,{\rm ms}$ after merger, most models show reduced instability, although some cases retain persistent unstable zones.

The regions of highest temperature roughly coincide with zones of convective stability, where, although there are some small unstable (red) regions, the Brunt-V\"ais\"al\"a is mostly positive (see Fig.~\ref{fig:tempvsbrunt}). In these regions, a fluid parcel that is vertically displaced will oscillate around its initial position rather than move farther away.  The buoyancy forces suppress convective motions, limiting the efficiency of heat redistribution by matter. As a result, thermal energy is not efficiently transported away and accumulates, leading to a sustained higher temperature. In contrast, in convectively unstable regions, heat is more easily redistributed by fluid motions, preventing such localized heating.
In Fig.~\ref{fig:tempvsbruntSLy}, we show the same correlation between high-temperature regions and convectively stable regions for a different EoS, SLy, with a lower magnetic-to-gas pressure.

It has also been suggested that there is a reverse causal relationship, in which a hotter region naturally becomes convectively stable~\cite{DePietri:2019mti}.
In such a scenario, increased thermal pressure in the hot region could balance buoyancy forces, suppressing convection, and supporting
a stable configuration. However, in the non-magnetized case, the high-temperature ring at $t-t_{\rm merg}=20\,{\rm ms}$ does not translate to a coincident region of convective stability $20\,{\rm ms}$ later (see first row of Fig.~\ref{fig:tempvsbrunt}). This region does not appear to influence the
long-term convective stability of the remnant, suggesting that temperature alone does not establish convective stability in this context.

On the other hand, we see that in the densest part of the star, where the temperature is the lowest, the values of the Brunt-V\"ais\"al\"a frequency are very close to zero, indicating that the matter is marginally buoyant. This is expected due to the small thermal and pressure gradients.

We remark the differences between the different magnetic field topologies. In particular, the pulsar-like case shows an extended and clearer stable region, with fewer red spots. Although its size and temperature profile are similar to that of configuration {\bf{T$_{0.5}$}}, the latter retains significant unstable regions, even at $40\,{\rm ms}$ after merger, consistent with sustained differential rotation, while the former pulsar-like case is almost rigidly rotating in the core (see Fig. \ref{fig:rot-WFF1}). If a fluid element moves into a region with a different angular velocity, centrifugal forces can enhance the displacement rather than restore it. Note that, for a differential rotating star, the criteria stability should account for differential rotation by also evaluating the epicyclic frequency, $\kappa^2=1/\varpi^3\nabla(\Omega^2\varpi^4)\cdot\nabla\varpi$, in which case we recover the Solberg-Høiland criteria \cite{Tassoul_2000}. At $40\,{\rm ms}$ after merger, cases {\bf{SP}} and {\bf{T$_{0.5}$}} retain a significant amount of differential rotation, while {\bf{T$_{0.95}$}}, although still differentially rotating, shows a less pronounced profile (see Fig.~\ref{fig:rot-WFF1}). If we were to account for the differential rotation, then we would expect more stability in the core of the star and less stability towards the surface.

In addition, the Brunt-V\"ais\"al\"a frequency provides information about the natural g-mode oscillation frequencies of the fluid in the remnant. In particular, it indicates that these frequencies are on the order of kilohertz in our systems. While for non-rotating or slowly rotating stars g-mode frequencies are typically $< 0.6 \,{\rm kHz}$ \cite{Zhao:2022tcw}, for rapidly rotating stars they can reach the kHz range \cite{Gaertig:2009rr}. The discrepancy between the frequency range obtained in our analysis and that reported in the literature is likely due to the approximations adopted in our approach (see Appendix~\ref{appendix:convective-stability}).

Finally, in Fig.~\ref{fig:tempvsbrunt} one can also notice how the magnetic field presence and topology impacts the structure and size of the remnants. As displayed in the third column, the non-magnetized case results in a more extended remnant. This is likely due to less efficient angular momentum redistribution in the absence of magnetic fields. In magnetized cases, effective magnetic viscosity triggers outward angular momentum transport and rapid dissipation of rotational energy, allowing the remnant to contract and become more compact. In contrast, in the non-magnetized case, angular momentum is redistributed only through numerical 
viscosity, which is a less effective mechanism for angular momentum transport. As a result, the remnant remains more extended and supported by differential rotation (see Fig.~\ref{fig:rot-WFF1}).
By contrast, the smallest remnants correspond to those with an initially extended poloidal component, cases $\mathbf{P}$ and $\mathbf{T}_{0.5}$. 

\begin{figure*}
    \centering
    \includegraphics[width=0.95\linewidth]{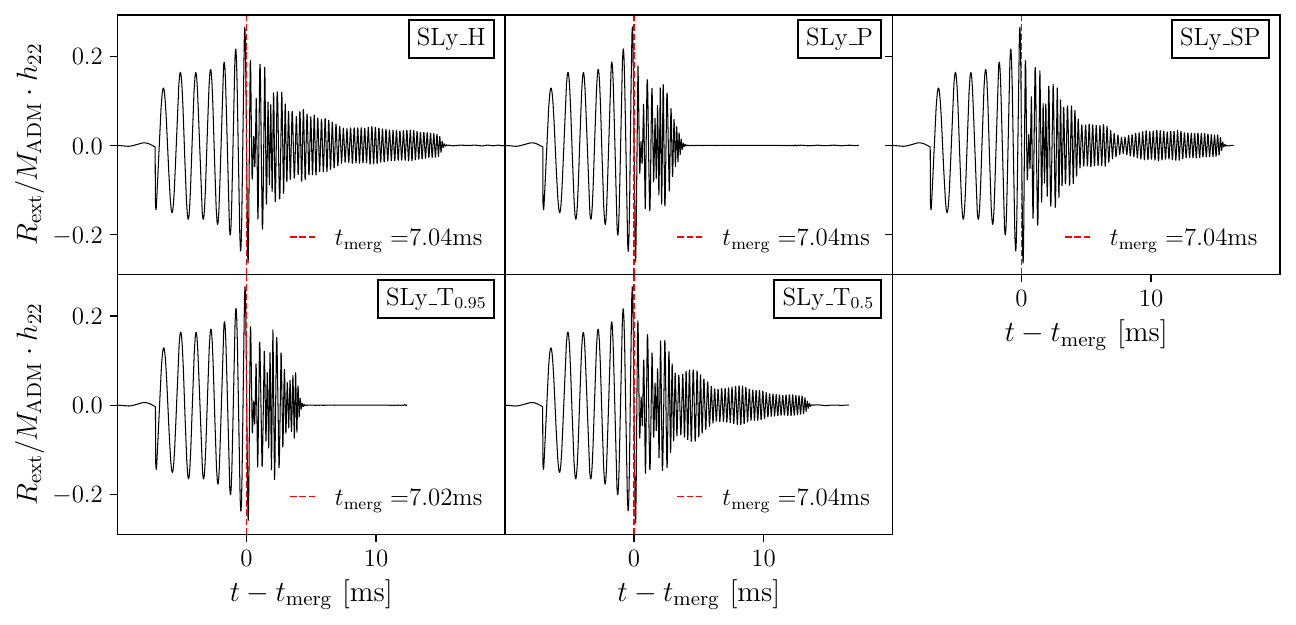}
    \caption{Gravitational wave strain of the dominant mode $h_{22}$ extracted at the coordinate radius $R_{\rm ext}\sim 1240\,{\rm km}\approx840M$ for the SLy EoS cases that undergo delayed collapse to a BH (see~Table~\ref{tab:results})  as a function of coordinate time measured from the merger. The merger time is defined as the peak of the GW signal that is indicated by the red dashed line.}
    \label{fig:SLy-waveforms}
\end{figure*}

\begin{figure*}
\centering
\includegraphics[scale=0.825]{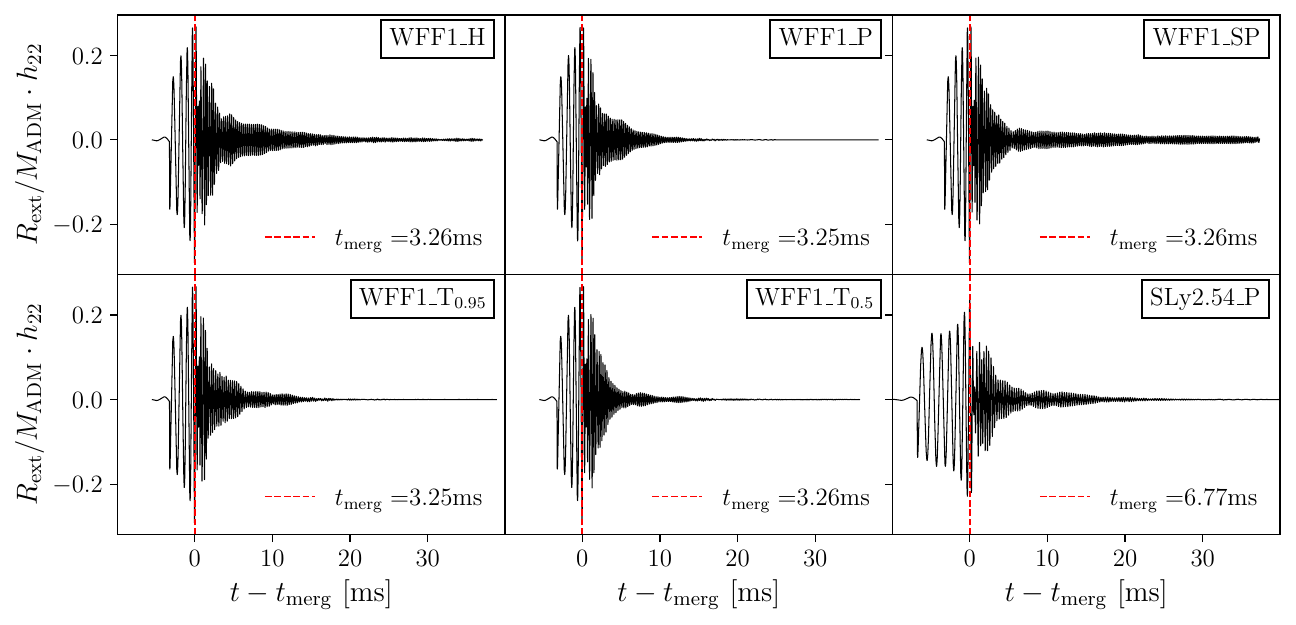}    
\caption{Gravitational wave strain of the dominant mode $h_{22}$ extracted at the coordinate radius $R_{\rm ext}\sim 620\,{\rm km}\approx420M$ for binary cases that for a long-lived remnant (WFF1 EoS and SLy2.54\_P in Table~\ref{tab:results})  as a function of  coordinate
    time measured from the merger (red dashed line).}
    \label{fig:waveforms}
\end{figure*}

\begin{figure*}
        \centering
    \includegraphics[scale=0.65]{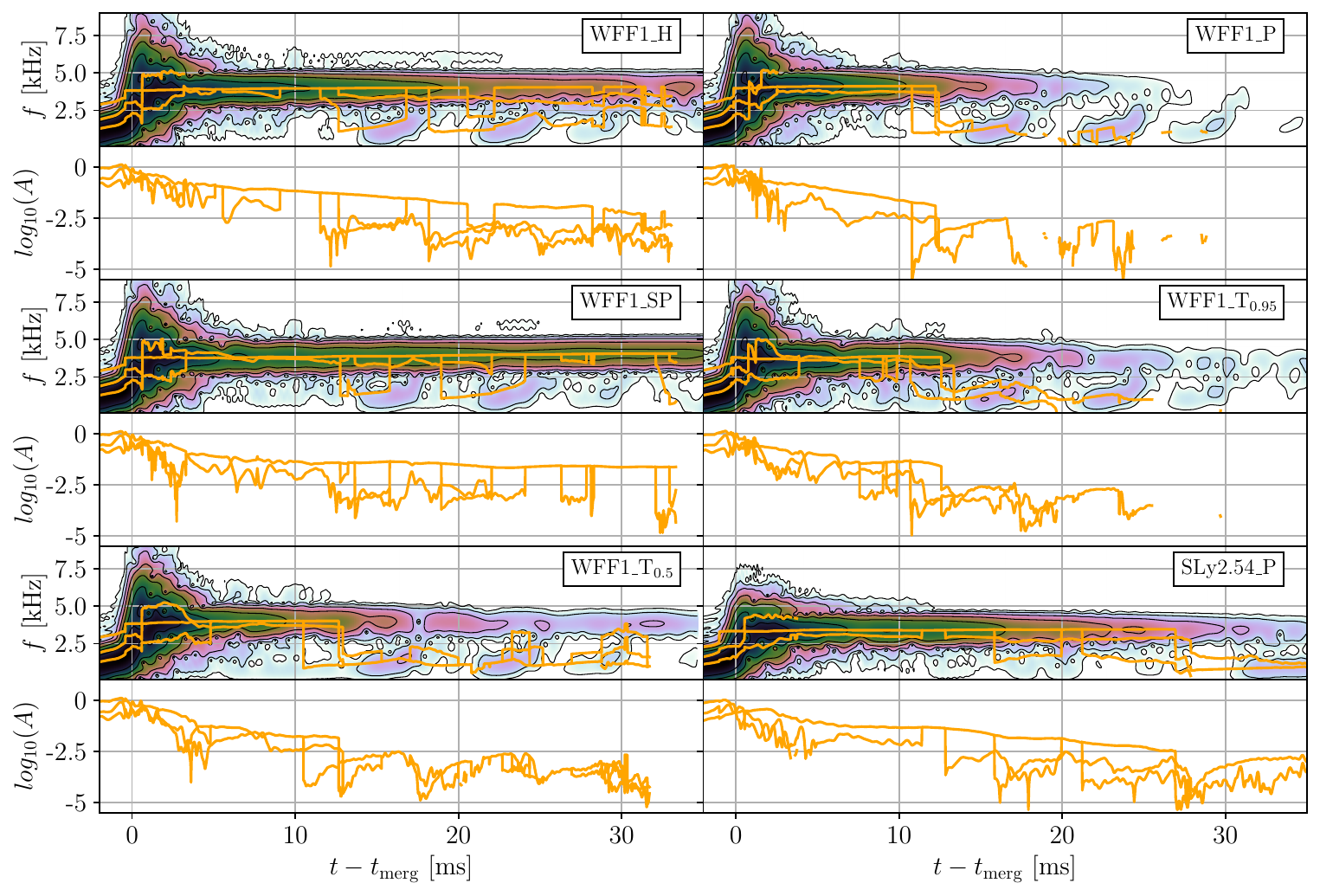}    
    \caption{The top half of each panel shows time-frequency spectrograms of the $l=m=2$ component of the GW strain for all models.
    Thick orange lines, computed from the Prony analysis, indicate the frequency of the active modes responsible for most of the GW emission at different times. Colors indicate the relative intensity of the spectral density (darker areas correspond to higher intensity). The bottom half of each panel displays the amplitude of the main active modes (in arbitrary units).}
    \label{fig:spectrograms}
\end{figure*}

\begin{figure*}
    \centering
    \includegraphics[scale=0.8]{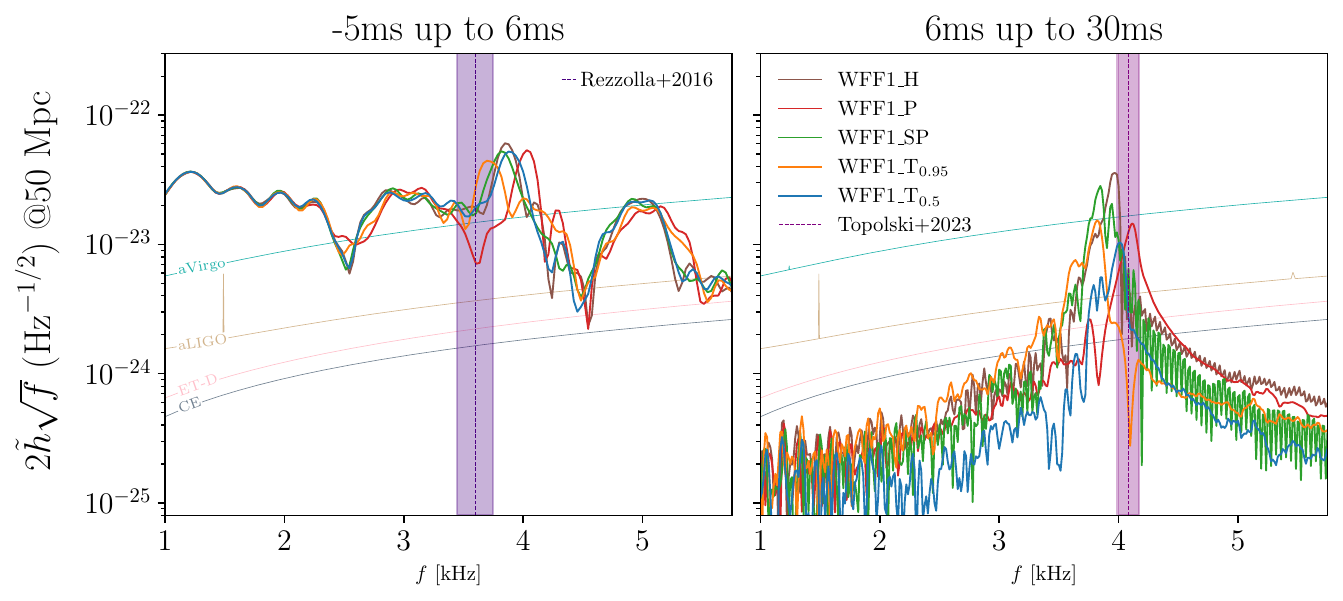}
    \caption{GW spectra for BNS mergers with WFF1 EoS at $50$ Mpc with optimal orientation. The spectra are shown for different restricted time windows (indicated at the top of each frame) in order to emphasize the contribution of the dominant spectral components at different times. For reference, we also show the sensitivity curves of Advanced Virgo \cite{Acernese_2015}, Advanced LIGO \cite{Aasi_2015}, Einstein Telescope~\cite{ET:2010}, and Cosmic Explorer \cite{Abbott_2017}. Vertical lines show the expected values for $f_{2i}$ and $f_2$ peaks on the left and right panels, respectively, from the quasi-universal relations from \cite{Rezzolla:2016nxn} and \cite{Topolski:2023ojc}. The shaded region around these lines covers 1$\sigma$.}
    \label{fig:fft}
\end{figure*}

\begin{figure*}
    \centering
    \includegraphics[width=0.95\linewidth]{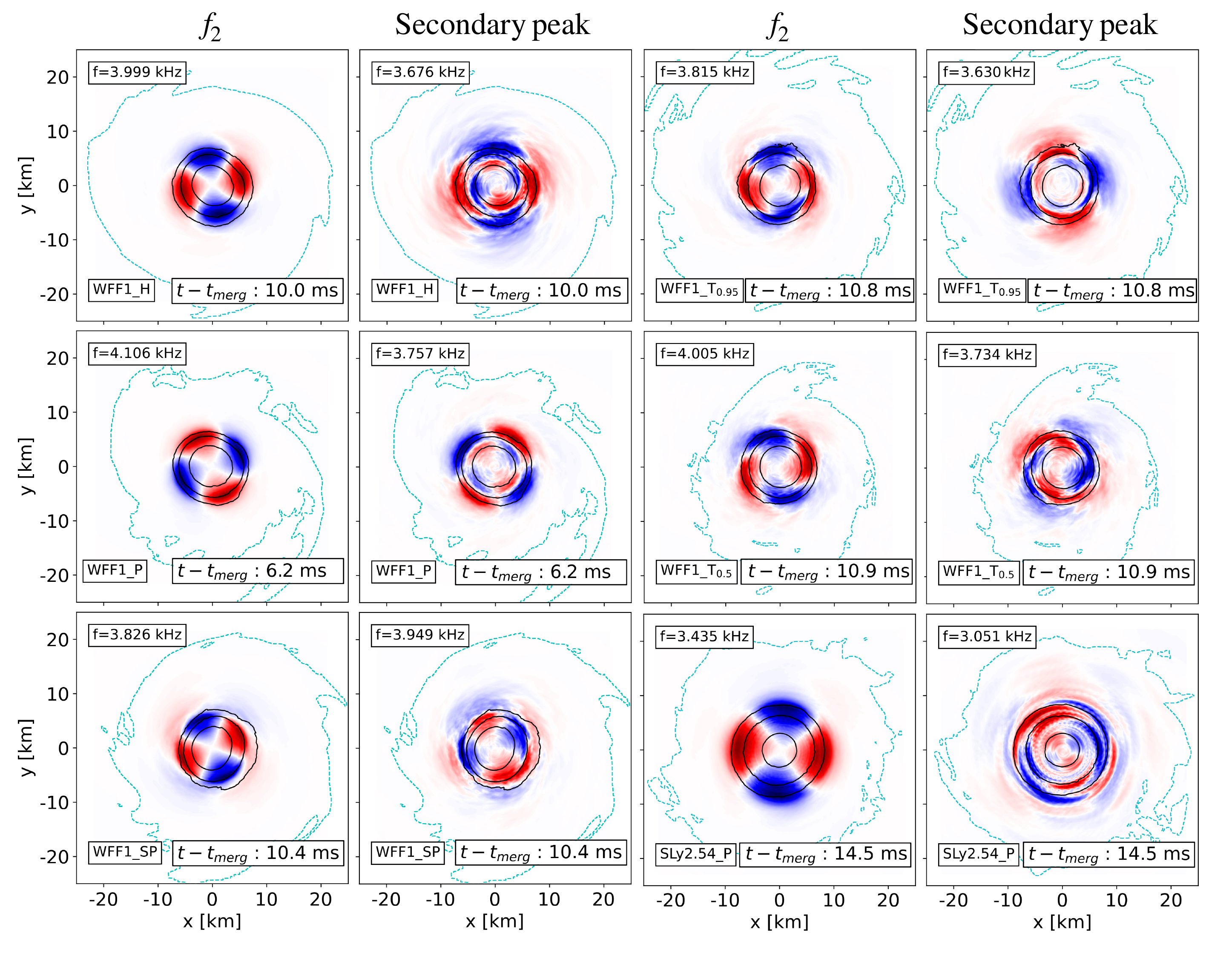}
    \vspace{-5mm}
\caption{Density eigenfunctions in the equatorial plane of  the WFF1 models in Table~\ref{tab:results}. The black lines are isocontours of the rest-mass  densities $1.76\times10^{14} \,{\rm g/cm}^3$, $5.01\times10^{14} \,{\rm g/cm}^3$, and $10^{15} \,{\rm g/cm}^3$ corresponding to the boundaries between segments of the piecewise equation of state used to model the star.
The cyan line indicates the surface of the star at $10^{-3}\rho_0^{\rm max}(t=0)$. The first and third columns display the $f_2$ mode, while the second and fourth show the additional mode observed 
in Fig.~\ref{fig:spectrograms}. The radial node in the density eigenfunctions of this mode suggests a coupling with the $m=0$ mode.}
    \label{fig:dens-eig-WFF1_P}
\end{figure*}

\begin{figure}
    \centering
    \includegraphics[width=\linewidth]{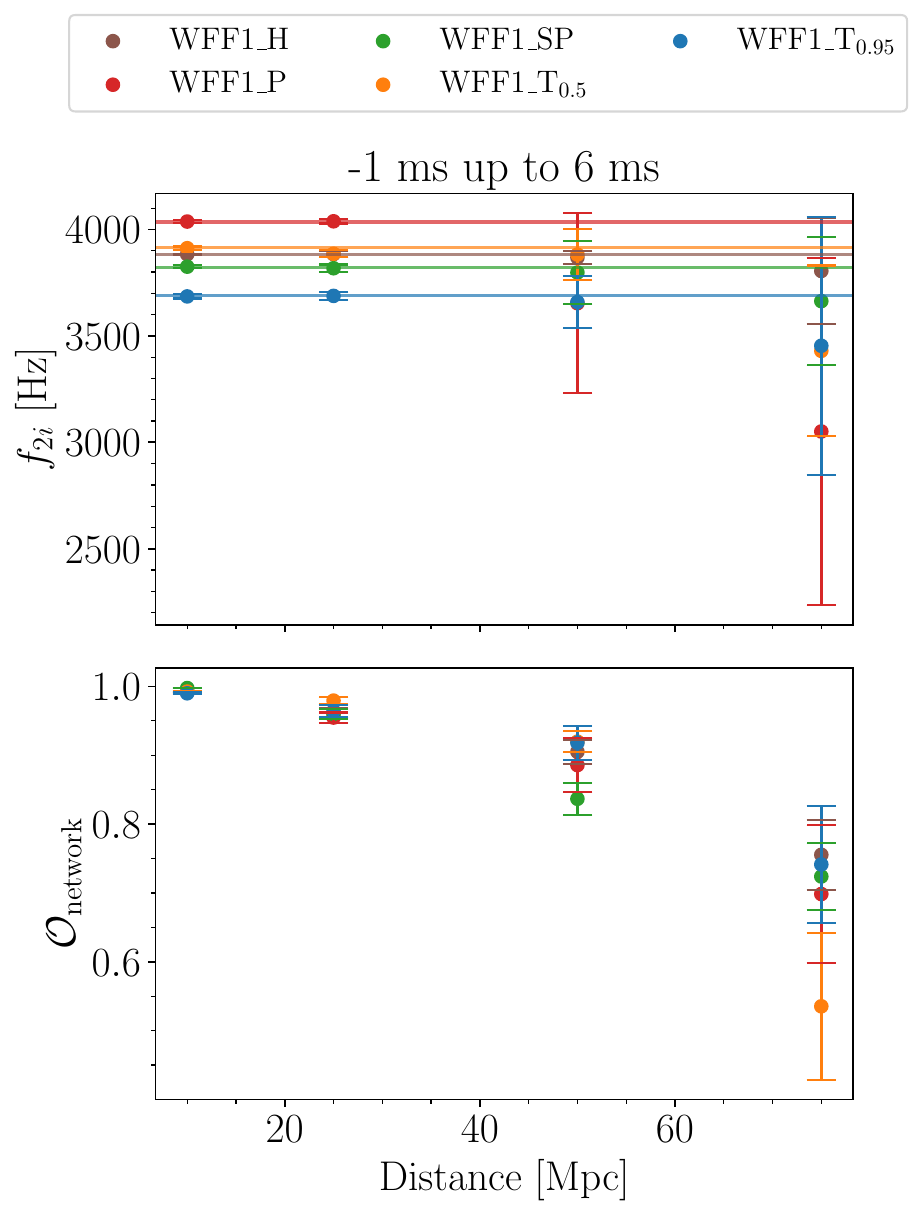}
\caption{Mean values (dots) of the recovered posterior distributions using the Einstein Telescope of the peak frequencies 
(upper panel) by the \textsc{BayesWave} algorithm and the network overlap (lower panel), together 
with their standard deviation (bars), as a function to the distance to the GW source. The injected
peak values are displayed by solid lines in the upper panel. The colors represent the different 
magnetic field topologies for the WFF1 configurations that do not undergo collapse to a BH.}
    \label{fig:fpeaks_ov_bw}
\end{figure}

\subsection{Gravitational waves}
Figs.~\ref{fig:SLy-waveforms} and~\ref{fig:waveforms} show the GW strain of the dominant mode $h_{22}$  
as a function of retarded time post-merger for the cases in Table~\ref{tab:results}. Before merger, the 
corresponding waveforms are nearly identical across all configurations, including both magnetized and 
non-magnetized evolutions.  This is expected, as the magnetic field is embedded in the stars two orbits prior 
to the merger.  We note that the merger time is the same in all the cases with the same EoS and initial magnetic-to-gas pressure.

Following merger, we observe a higher-frequency signal arising from oscillations of the newly bar-deformed 
binary remnant. In the HMNS SLy cases, depending on the magnetic field content, and thus the effective magnetic 
viscosity, the post-merger GW signal decreases in amplitude and either: i) comes to an abrupt end following the 
quasinormal ringdown modes of the BH, as seen in the {\bf{P}} and  ${\bf{T}}_{0.95}$ configurations (cases with 
stronger magnetic field; see Fig.~\ref{fig:SLy-waveforms}), within a timescale of $t - t_{\rm merg} \sim 3\,{\rm ms}$ 
(see Table~\ref{tab:results}  for details); or ii) follows an oscillatory behavior with a slowly varying envelope, 
as in the {\bf{H}}, {\bf{SP}} and {\bf{T$_{0.5}$}} configurations, which terminates when the remnant 
collapses to a BH at $t - t_{\rm merg} \sim 15\,{\rm ms}$.

On the other hand, the waveforms associated with the WFF1 binaries show similar behavior during the early post-merger phase, with significant differences only emerging after $t-t_{\rm merg}\gtrsim 5\,{\rm ms}$ (see Fig.~\ref{fig:waveforms}). In the non-magnetized (WFF1\_H) and 
superposition (WFF1\_SP) cases, the quadrupolar oscillation modes of the fluid persist throughout the duration of the simulation, 
resulting in a sustained GW signal with little post-merger damping. We note that the GW amplitude of WFF1\_SP is notably larger than that of WFF1\_H at late times. One might instead expect the magnetic field to enhance angular momentum redistribution, thereby driving the remnant more rapidly toward axisymmetry and damping the GW emission more efficiently in the magnetized case. A possible explanation for this behavior is that the frozen-in magnetic field effectively makes the initially ellipsoidal remnant more resistant to deformation, due to the additional magnetic pressure support. As a result, the non-axisymmetric, ellipsoidal structure can persist for a longer time, sustaining GW emission. Furthermore, since the MRI is not fully resolved in this configuration, the associated angular momentum transport is likely underestimated and therefore insufficient to counteract this effect. In contrast, configurations with strong magnetic fields (i.e. WFF1\_P, WFF1\_T$_{0.95}$, and WFF1\_T$_{0.5}$) show rapid damping of the waveform amplitude, becoming negligible by $t-t_{\rm merg}\gtrsim 15\,{\rm ms}$. This suppression of post-merger oscillations indicates the formation of a uniformly rotating core, driven, in particular, by magnetic braking. This interpretation is consistent with the rotational profiles (see~Fig.~\ref{fig:rot-WFF1}), which show a magnetically braked core surrounded by a low-density, Keplerian cloud.

\paragraph{\bf GW Spectrum and Prony's analysis for longer-lived remnants:}
To further analyze the behavior of the stable remnants, we study the time-frequency representation of the 
main GW mode. The general behavior of our six longer-lived systems is effectively illustrated by the spectrogram in
Fig.\ref{fig:spectrograms}, which also includes thick (orange) lines that trace the time evolution of the dominant 
spectral modes of the remnant. These frequencies are estimated using the {\tt {ESPRIT}} Prony's method with 
a $4\,{\rm ms}$ moving window (see Sec.~\ref{sec:diagnostics}). The corresponding mode amplitudes, presented in 
arbitrary units, are shown in the bottom part of each panel. The relative intensity of the spectral density 
is also displayed, with darker areas corresponding to higher intensity. We note  the spectrograms show a 
higher intensity shortly after merger, and not exactly at $t = t_{\rm merg}$, as the analysis employs a moving
window that captures preceding dynamics.

As the magnetic field during the inspiral epoch is only advected with the fluid, and so does not change the system,
we observe that at merger the instantaneous frequency $f_{\rm max}$  across the WFF1 cases is $2\,{\rm kHz}$. However, 
during merger magnetic instabilities are triggered, leading to a significant amplification of the magnetic field. 
During this epoch, all models exhibit very similar spectrograms in both  frequency and time. The dominant transient 
$f_{2i}$ frequency is $\sim 4.0\,{\rm kHz}$ for WFF1, which coincides with twice the rotation frequency of  the bar 
deformation of the binary remnant (see Fig.~\ref{fig:rot-WFF1}). However, we note a change in frequency of 
the order of $\pm 200\,{\rm Hz}$  depending on the magnetic field configuration (see Table~\ref{tab:results}), which
remains throughout the post-merger window, i.e. this shift in frequency is also evident in the $f_2$ frequency.
As shown in the bottom of each panel in Fig.~\ref{fig:spectrograms}, the amplitude of the dominant mode decreases
over time, and is fully damped in SLy2.54\_P and in both WFF1\_P and WFF1\_T$_{0.95}$ by $t-t_{\rm merg}\sim 25\,{\rm ms}$. Similar 
shifts in frequency due to magnetic viscosity have been reported in~\cite{Bamber:2024qzi,Tsokaros:2024wgb}.

Another feature similar in all the spectrograms, independently of their magnetic field topology, is the excitation 
of an intermittently low-frequency mode of around  $\sim1.25\,{\rm kHz}$, which may be linked to either the reduction 
in differential rotation (see Fig.~\ref{fig:rot-WFF1}), or the gradual decay of the perturbation amplitude as shown 
in the bottom  of each panel in Fig~\ref{fig:spectrograms}. Note that the time at which this mode is excited depends 
on the magnetic field configuration, and also decreases over time. Additionally, there is a third mode closer to the 
$f_2$ mode  that is also  active in all cases, although it stands out the most from $f_2$ in the {\bf{P}} case.  The fact that 
the frequency of this mode is close to the frequency of the $m=2$ mode suggests that this is not one of its overtones
(see below).

\paragraph{\bf Power spectral density (PSD):}
Figure~\ref{fig:fft} shows the power spectral density (PSD) of the GW signal for  WFF1 cases computed using  two distinct 
post-merger time windows. As mentioned before, the magnetic field influences the peak frequency of the dominant  mode in 
both the early and late post-merger phases. It is worth mentioning that the  {\bf P} model displays the highest peak frequency  
with $f_2\simeq 4.11\,{\rm kHz}$.

On the other hand,  WFF1\_SP and WFF1\_T$_{0.95}$ models have lower peak frequencies than that of the non-magnetized case. 
As the $f_2$ can be correlated with the compactness of the remnant,  more compact remnants  tend to have higher $f_2$ 
values, while less compact ones yield lower frequencies~\cite{Takami:2014tva,Rezzolla:2016nxn}. While this trend is observed 
for the magnetized cases,  it is noteworthy that the non-magnetized remnant is actually the largest in size (see, e.g. 
Fig.~\ref{fig:tempvsbrunt}, where the bulk of the star is outlined by a green dashed line). Thus, one would naively expect 
this case to produce the lowest $f_2$  frequency. The fact that configurations {\bf SP} and {\bf T}$_{0.95}$ exhibit even 
lower $f_2$ values suggests that the peak frequency is influenced by other physical properties of the star remnant.

Along with the power spectral density, we also plot vertical lines corresponding to the expected values for $f_{2i}$ and $f_2$ 
peaks from the quasi-universal relations from~\cite{Rezzolla:2016nxn,Topolski:2023ojc}. We observe that both of these quasi-universal 
relations deviate for this equation of state, and that the $1\sigma$ interval  fails to account for the frequency shifts 
caused by differing magnetic field topologies.   
Notice that, as pointed out in~\cite{Tsokaros:2024wgb}, non-magnetized BNSs suggested that shifts in the main post-merger GW 
peak could be the smoking gun to feature some properties of the EoS like phase transitions. However, our results suggest that magnetic fields alone can produce similar frequency shifts, as reported in \cite{Tsokaros:2024wgb,Bamber:2024qzi}.

\paragraph{\bf Density eigenfunctions:}
To study the main oscillation modes present in the remnants, whose corresponding frequencies are shown in Fig.~\ref{fig:spectrograms} and identified as the peaks in Fig.~\ref{fig:fft}, we performed a FFT over $\sim 2\,{\rm ms}$ intervals of the density distribution and identified the Fourier amplitude at selected fixed frequencies, corresponding to the eigenfunction of a particular mode as done in \cite{Stergioulas:2003ep, Stergioulas:2011gd, DePietri:2019mti}. We labeled each segment using its midpoint time.

Fig.~\ref{fig:dens-eig-WFF1_P} shows the density eigenfunctions at the frequencies identified in the spectrograms as the main active modes, at a given time post-merger. The time displayed is different for each remnant, and is chosen based on the clearest images obtained from this analysis. Each panel is scaled to the peak value of the density eigenfunction presented. 

The first and third columns display the eigenfunction of the $f_2$ mode for each WFF1 configuration in the equatorial plane. A telltale sign that this corresponds to the $f_2$ mode is its quadrupolar shape. The absence of radial nodes in this plane suggests that this mode is the fundamental one.  As mentioned before, using Prony's analysis we identify a new mode with a frequency close to the fundamental one (see Fig.~\ref{fig:spectrograms}). The density eigenfunctions for this frequency are shown in the second and fourth columns. We notice WFF1\_P displays a clear node along the radial direction. We exclude the possibility of this being an overtone of the $m=2$ mode, since its frequency is very close to the corresponding $f_2\approx4.11\,{\rm kHz}$. However, the radial node is also a characteristic of the $m=0$ mode (see e.g the top panel of Fig.~1 in \cite{Stergioulas:2011gd}). The frequency of the $m=0$ mode can be extracted from the oscillations of the lapse, and it is $\sim 1\,{\rm kHz}$. Although the frequency of this additional mode is not exactly $f_2-f_0$ or $f_2+f_0$, we can conclude this is a non-linear coupling of modes $m=0$ and $m=2$ due to the topology of its density eigenfunctions, which exhibit both the quadrupolar shape of the $m=2$ mode and the single radial node of the $m=0$ mode. Although the additional mode of the other remnants display a similar structure, their nature is not as clear as in WFF1\_P. One particular thing to look out for is the decreasing amplitude in the radial direction, which can indicate that the presence of a radial node is due to differential rotation and does not correspond to a new mode \cite{Stergioulas:2003ep, DePietri:2019mti}. Another indicator of this is that the node is farther away from the center of the star. An obvious example of this is WFF1\_SP, which retains a high degree of differential rotation even at $40\,{\rm ms}$ after merger (see Fig.~\ref{fig:rot-WFF1}). Therefore, we can only conclude with confidence that, during the time of our evolutions, the pulsar-like configuration excites a non-linear coupling of the $m=0$ and $m=2$ modes.

\paragraph{\bf Detectability of the shifts in the peak frequency:}
To assess the detectability of the shifts in the peak frequency arising from different magnetic field topologies, 
we use the {\tt BayesWave} algorithm~\cite{Cornish:2015}, which reconstructs unmodeled GW signals using a basis 
of sine-Gaussian wavelets and minimal assumptions~\cite{BECSY:2017}. The algorithm employs a trans-dimensional 
reversible jump Markov Chain Monte Carlo (RJMCMC) to jointly sample the posterior distributions of the wavelet 
parameters. From these, posterior distributions for derived quantities, such as the GW spectra discussed above, 
are obtained by applying a FFT to the reconstructed signals, with a boxcar (rectangular) window. This allows us to compute 
posteriors for the peak frequencies associated with the dominant post-merger modes (see discussion in 
Sec.~\ref{subsec:post-merger_evol}). The analysis is performed using time-domain waveforms from WFF1 models,  
those that do not  undergo collapse, injected into Gaussian noise consistent with the sensitivity of the Einstein 
Telescope (ET).  We adopt the ET-D configuration~\cite{Hild:2011}, consisting of a three-detector network at a single 
site, and  assume optimal sky location and inclination. For the wavelet functions, we set a maximum number of wavelets of $N_{\rm max} = 200$ and a maximum quality factor of $Q_{\rm max} = 200$. We also employ $n = 4 \times 10^6$ iterations, and a sampling rate of $16384\, {\rm Hz}$. We use uniform offset phase parameters in the range $\phi_0 =[0,2\pi]$, and the signal
wavelet amplitude prior is presented and discussed in~\cite{Cornish:2014kda}. This setup enables us to determine the potential detectability 
of frequency shifts induced by varying magnetic topologies, particularly the $f_2$ mode and additional subdominant modes 
identified in the spectrograms.

The upper panel in Fig.~\ref{fig:fpeaks_ov_bw} shows the mean value of the recovered peak frequencies, together 
with their standard deviation, as a function of the distance to the GW source. The horizontal lines represent 
the injected values. The time window used to compute the GW spectra is chosen to capture the early post-merger 
phase, i.e. from $t-t_{\rm merg} = -1\,{\rm ms}$ to $t-t_{\rm merg} = 6\,{\rm ms}$, where the signal amplitude is larger. 
For  distances up to $\sim 25$ Mpc, all the frequency peaks are distinguishable. At a distance of $D = 50\,{\rm Mpc}$, 
most peaks are still distinguishable, but the one from WFF1\_P shows a larger discrepancy, due to the peak being located at higher frequencies, where the detector noise is greater. At a distance of $75\,{\rm Mpc}$, the recovered values overlap, and thus these shifts may no longer be detectable at this distance.

A way to check the reconstruction performance of the algorithm is by computing the overlap function between the injected and recovered signals. This function can take values from -1 to +1. A perfect match between the signals will result in $\mathcal{O} = +1$, and a perfect anti-correlation will give $\mathcal{O} = -1$. If there is no similarity between the signals, the overlap will be 0. The expression for the weighted overlap of a network of $N$ detectors is~\cite{Cornish:2014kda}
\begin{equation}\label{net_overlap}
    \mathcal{O}_{\rm network} = \frac{\sum^N_{k=1}\langle h_i^{(k)}\,, 
    h_r^{(k)} \rangle}{\sqrt{\sum^N_{k=1}\langle h_i^{(k)},h_i^{(k)}
    \rangle}\sqrt{\sum^N_{k=1}\langle h_r^{(k)},h_r^{(k)}\rangle}}\,,
\end{equation}
where $h_i$ and $h_r$ denote the injected and recovered signals, respectively, and the index $k$ refers 
to the $k$-th detector.  This expression involves the inner product of two complex quantities, which can 
be defined as 
\begin{equation}\label{inner_prod}
    \langle a,b \rangle \equiv 4 {\rm Re}\left[\int_{f_{\rm min}}^{f_{\rm max}}\frac{a(f)b^*(f)}{S_h(f)}df\right]\,,
\end{equation}
where $S_h(f)$ refers to the noise PSD of the detector and $(f_{\rm min}, f_{\rm max}) = (2000,6000)$ is the 
bandwidth of the analysis. Note that we only consider the post-merger frequencies of interest. Here, the 
asterisk indicates complex conjugation.

The lower panel of Fig.~\ref{fig:fpeaks_ov_bw} shows that up to $\sim 25\,{\rm Mpc}$, the network overlap is roughly 
$1$, implying that the injected and recovered signals match almost perfectly. However, as the distance to the 
source increases, the overlap drops to lower values. At  $D = 50\,{\rm Mpc}$ , the overlap lies between $\sim 0.8-0.9$. 
The match is still large, but since the frequency peaks are very close for the different magnetic field configurations, 
the recovered peaks at this distance already overlap in some cases. For a distance of $D = 75\,{\rm Mpc}$, the network 
overlap falls below $0.8$, in agreement with the large uncertainty in the recovered peak frequencies. Therefore, we conclude that shifts in the early $f_{2i}$ frequency peaks arising from different magnetic field configurations could be detected by third-generation detectors, e.g. Einstein Telescope, for distances up to $\approx 50$ Mpc.

\begin{figure}
    \centering
    \includegraphics[width=\linewidth]{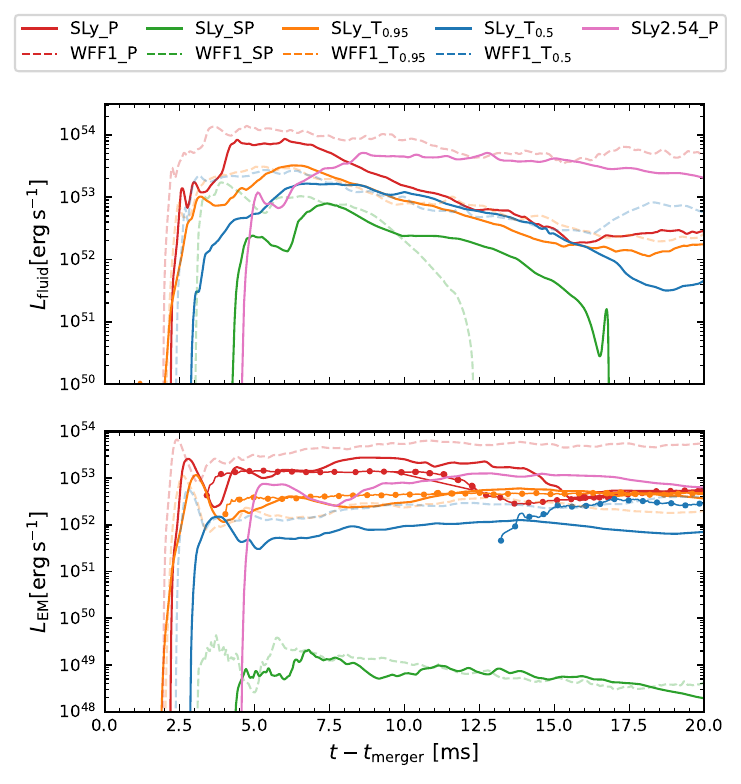}
    \caption{Luminosity via fluid outflow (top) as computed by Eq.~\eqref{eq:fluid-luminosity} 
    and EM radiation (bottom) computed by Eq.~\eqref{eq:poynting-luminosity} as a function of 
    the coordinate time since $t_{\rm merg}$ (defined as the time of peak GW emission) extracted 
    over a sphere at $r_{\rm ext} = 539\,\rm km$. The dots and corresponding connecting lines in the 
    Poynting luminosity plot display the expected BZ electromagnetic luminosity from Eq.~\eqref{eq:BZ-lum} 
    for the BH mass, spin and approximate $B_{\rm pol}$ for cases leading to a BH.}
    \label{fig:luminosity}
\end{figure}

\begin{figure*}
\includegraphics[width=0.85\linewidth]{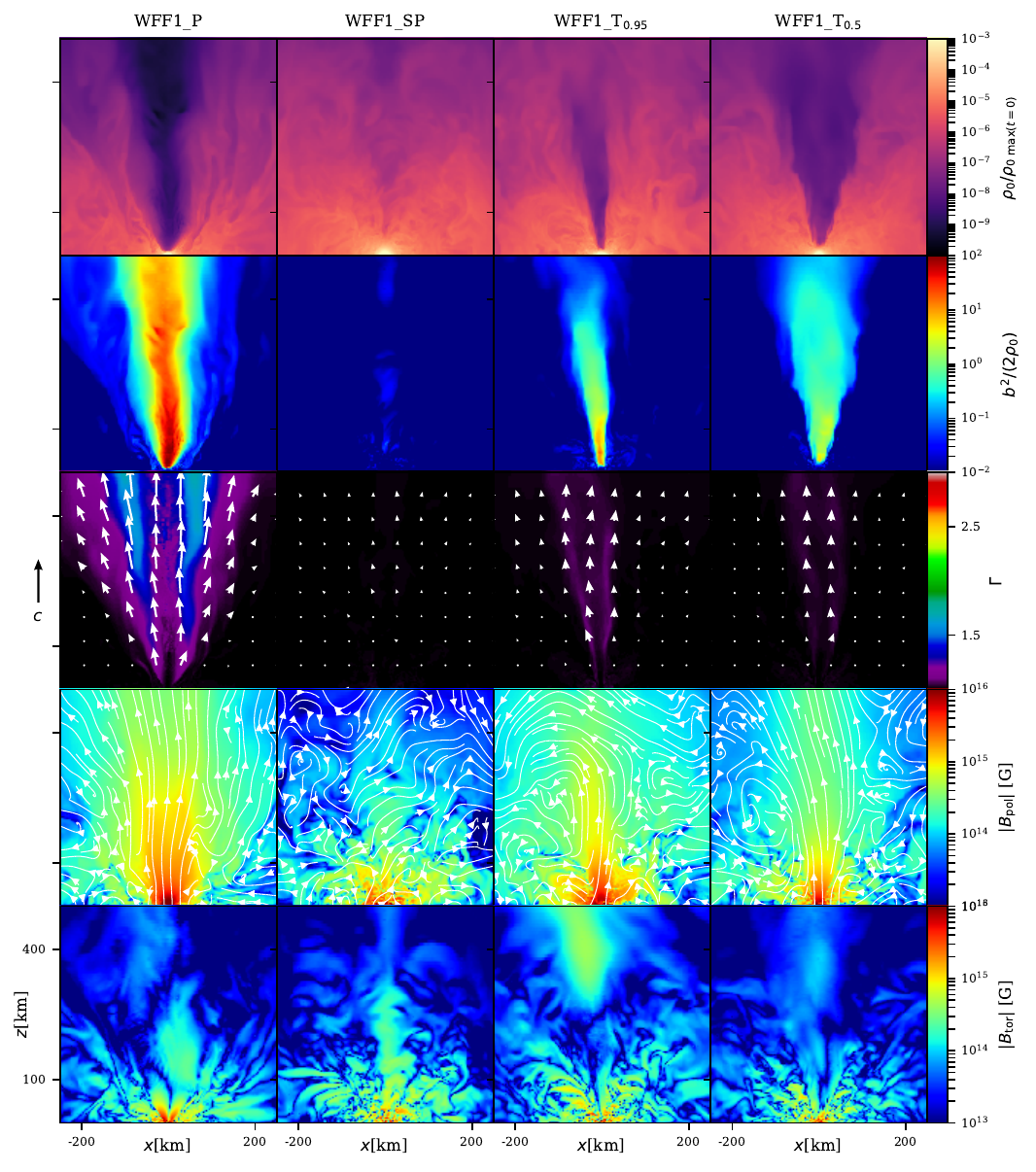}
    \caption{Meridional slides at $t-t_{\rm merg}\sim 40\,{\rm ms}$ for the magnetized WFF1 BNS remnants in Table~\ref{tab:results}. We show the rest-mass density 
    $\rho_0$ on a logaritmic scale relative to the initial maximum rest-mass density (top row), $b^2/2\rho_0$ which approximately corresponds to the magnetization $\sigma$ (second row), the Lorentz factor (third row), and the strength of the poloidal and toroidal magnetic field (fourth and fifth rows). White lines indicate the magnetic field lines while the arrows indicate the flow velocities. The arrow labeled on the left-hand side indicates the magnitude of the speed of light.}
    \label{fig:b2o2rho}
\end{figure*}

\subsection{Electromagnetic signals}
\label{seb:EM_signal}
\subsubsection{Consistency with sGRB}
The fluid and Poynting luminosities, as defined by Eqs.~\eqref{eq:fluid-luminosity} and~\eqref{eq:poynting-luminosity}, respectively, are shown in Fig.~\ref{fig:luminosity} for all the cases. We notice that for both EoS the Poynting luminosity of the {\bf{SP}} configuration is significantly lower than that of the other cases, with values $\sim 10^{48}- 10^{49} {\rm erg\, s}^{-1}$, while the other configurations range between $\sim 10^{51}- 10^{53} {\rm erg\, s}^{-1}$.

The electromagnetic luminosity computed from our simulations allows us to estimate the isotropic-equivalent $\gamma$-ray luminosity of a jet observed head-on via \cite{Bamber:2024kfb}
\begin{equation}
    L_{\gamma,\rm iso}=\frac{1}{C_{\rm col}}\eta_{\rm EM}L_{EM} \, ,
\end{equation}
where $C_{\rm col}$ accounts for $\gamma$-ray collimation \cite{Khan:2018ejm} and is $\sim10^{-1}-10^{-2}$ for perfect collimation and half-opening angles of $10^\circ-30^\circ$. $\eta_{EM}$ is an efficiency factor, whose value is highly uncertain as the mechanism responsible for the prompt emission is still not yet fully understood. Nevertheless, $10^{-1}$ has been taken as a fiducial value for outflows from a black hole central engine \cite{Khan:2018ejm}. We expect that, for the remnant neutron stars, this factor will be smaller, $\sim10^{-2}$. The estimated isotropic-equivalent $\gamma$-ray luminosity of the {\bf{SP}} configurations for perfect collimation lies in the range $\sim10^{48}-10^{49}{\rm erg\, s}^{-1}$, partially consistent with observed sGRB. However, as we discuss below, this configuration does not exhibit collimation, and there is no evacuated funnel. On the other hand, the remaining simulations yield $\sim10^{51}-10^{54}{\rm erg\, s}^{-1}$, consistent with observed sGRBs \cite{Li:2016pes, Beniamini:2020adb, Shapiro:2017cny}.

The lack of a magnetically-dominated funnel explains the fact that the fluid luminosity becomes negative for the {\bf{SP}} cases around $17$ and $12.5\,{\rm ms}$ after merger for the SLy and WFF1 EoS, respectively, since there is no outflow.

We note that both the fluid and Poynting luminosity of the configuration {\bf{T$_{0.95}$}} are largely independent of the EoS (compare the solid and dashed orange lines), while case {\bf{P}} shows the largest difference between the two EoS (compare the solid and dashed red lines). This difference in EoS dependence can be explained by the different outcomes of the simulations, both in terms of the fate of the remnant and the magnetization levels. The difference observed between cases SLy\_P and WFF1\_P is likely due to the accretion of material onto the BH in the former. This hypothesis is supported by the fact that SLy2.54\_P, which does not collapse, presents fluid luminosity values of the same order of magnitude as WFF1\_P. One could expect the same to be true for the SLy\_T$_{0.95}$ and WFF1\_T$_{0.95}$, as the fate of their remnants is the same as their {\bf{P}} counterparts. However, their magnetization levels are different. The second row of Fig.~\ref{fig:b2o2rho} shows the magnetization levels of the WFF1 cases at $\sim40\,{\rm ms}$ after the merger, where we see that the {\bf{P}} configuration presents much larger magnetization values than the ones of {\bf{T$_{0.95}$}}, which translates to an increased particle acceleration. The poor acceleration in the {\bf{T$_{0.95}$}} configuration leads to less amounts of unbound material (see Fig.~\ref{fig:ejecta-mass}) even if no BH was formed, and therefore, for this magnetic field configuration, the accretion onto the BH does not have a big impact on the fluid and Poynting luminosities, nor on the ejecta mass.

Unlike configuration {\bf{T$_{0.95}$}}, and despite showing similar ejecta masses (see Fig.~\ref{fig:ejecta-mass}), cases with {\bf{T$_{0.5}$}} yield, throughout the evolution, Poynting luminosities of different orders of magnitude for each EoS. Notice that SLy\_T$_{0.5}$ and WFF1\_T$_{0.5}$ exhibit similar fluid luminosities, especially in the range $t-t_{\rm merg} \sim 6.5 -15\,{\rm ms}$, diverging only after SLy\_T$_{0.5}$ collapses at $t-t_{\rm merg}\approx13.17\,{\rm ms}$ (the time difference can be explained by the fact that the fluid luminosity is extracted at $r_{\rm ext}=539\,{\rm km}$, and so the effects of the collapse can only be noticed in the fluid luminosity at later times). In particular, when SLy\_T$_{0.5}$ forms a BH, the magnetic energy drops (see Fig.~\ref{fig:magnetic-energy}), suggesting that the acceleration of particles will be reduced for this case. However, the difference in the Poynting luminosity is roughly constant and is probably attributed to both the differences in the ejected mass as a result of the different EoS and the slight differences in the magnetic energy of these two cases.

The lower panel of Fig.~\ref{fig:luminosity} also displays dots with connecting lines that correspond to the expected BZ electromagnetic luminosity as computed via Eq.~\eqref{eq:BZ-lum}. The expected BZ luminosity for the {\bf{SP}} case is below the lower limit of the y-axis range and therefore is not seen in the plot. The other three cases show good agreement between each other by the end of our simulations, in particular cases {\bf{P}} and {\bf{T$_{0.95}$}} which collapse around the same time and whose BHs have the same mass and spin (see Table~\ref{tab:results}). The Poynting luminosity of the {\bf{T$_{0.95}$}} configuration agrees for the most part with the one expected from the BZ mechanism, while the pulsar-like configuration differs from the expected one around $t-t_{\rm merg}\sim 7.5-15\,{\rm ms}$, but their agreement by the end of the simulation indicates that the BZ mechanism dominates then. In contrast, the Poynting luminosity of {\bf{T$_{0.5}$}} does not match the BZ one. This is likely related with the ``late"-time formation of a BH in that case, which might not have allowed enough time for the BZ mechanism to become dominant by the time we terminated this simulation.

The Poynting flux generated by the rotating coiled magnetic field accelerates an outflow wind of gas, forming an evacuated funnel (see top row of Fig.~\ref{fig:b2o2rho}). Naturally, this does not happen for the {\bf{SP}} configuration, as the Poynting luminosity is significantly lower (there is no jet). We also note that, for the {\bf{P}} configuration, the funnel has a significantly lower density than the other cases. Between configurations {\bf{T$_{0.95}$}} and {\bf{T$_{0.5}$}}, the latter presents lower densities in a broader region, suggesting that a poloidal component might play a key role in evacuating the funnel.
In the second row of Fig.~\ref{fig:b2o2rho}, we show $b^2/(2\rho_0)$, which approximately corresponds to the magnetization, and is closely related to the terminal Lorentz factor $\Gamma_\infty$ (see Eq.~\eqref{eq:Lorentz_infty}), therefore being a good proxy for sGRB-consistent incipient jets. This is naturally connected to the first row, as lower density regions will present larger magnetization values. Thus, the pulsar-like configuration is the only one with $\Gamma_\infty$ consistent with sGRBs ($\sim10^2$). While {\bf{T$_{0.95}$}} reaches large $\Gamma_\infty$ values, this is restricted to the funnel region close to the remnant. With smaller values of $b^2/(2\rho_0)$, {\bf{T$_{0.5}$}} presents larger values than {\bf{T$_{0.95}$}} farther away, consistent with its clearer funnel. The third row of Fig.~\ref{fig:b2o2rho} shows that the {\bf{P}} case accelerates the fluid more efficiently, reaching Lorentz factors within the simulation box of $>1.5$, whereas case {\bf{T$_{0.95}$}} reaches at most $\Gamma\sim1.3$, and the other cases presenting even smaller values. The larger values in case {\bf{T$_{0.95}$}}, when compared to {\bf{T$_{0.5}$}}, can be explained by poloidal magnetic field strength in these cases (see the fourth row of Fig.~\ref{fig:b2o2rho}). While {\bf{T$_{0.5}$}}'s magnetic field lines seem to be more collimated, the poloidal component is stronger in {\bf{T$_{0.95}$}}, leading to larger Lorentz factors.

We note that, although case {\bf{T$_{0.95}$}} starts with an exterior poloidal field, its ``late"-time exterior topology differs significantly from that of case {\bf{P}}. This is not unexpected as most of the magnetic energy is concentrated in the stars. However, we do note that there is some degree of collimation, especially until $\sim350\,{\rm km}$ above the equatorial plane (see Fig.~\ref{fig:visit-3dplot-SLy-case1} and the fourth row of Fig.~\ref{fig:b2o2rho}).

On the other hand, the {\bf{SP}} case shows no degree of collimation, and the strength of the poloidal component of its magnetic field is considerably lower than that of the other cases, which we already had observed upon magnetic field insertion (see Table~\ref{table: initial magnetic field}) and also expected as we do not completely capture the MRI.

Consistent with Fig.~\ref{fig:magnetic-energy}, the pulsar-like configuration presents both the strongest poloidal and toroidal magnetic field components. We conclude that the only topology (from the ones tested) compatible with a highly collimated, relativistic outflow of matter consistent with sGRBs is a purely poloidal one. 

\subsubsection{Ejecta mass and kilonova estimates}
\label{sec:ejecta}
\begin{figure}
    \centering
    \includegraphics[width=\linewidth]{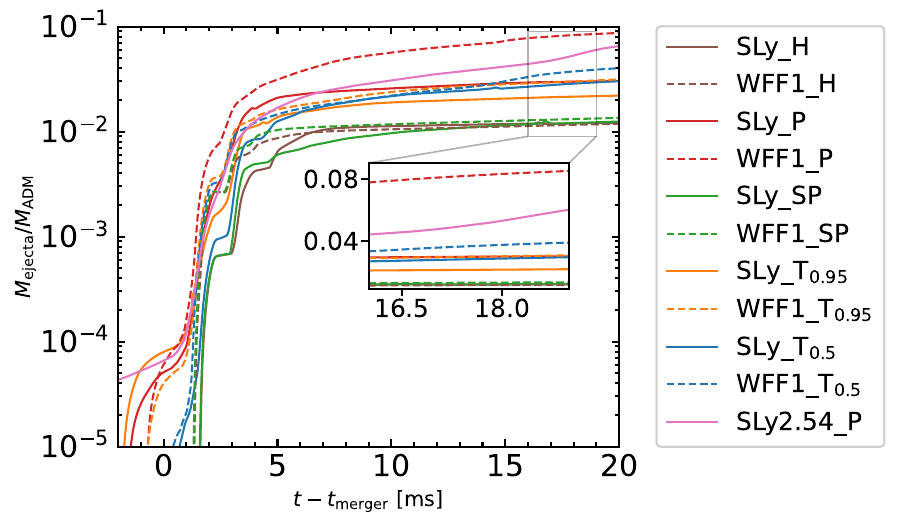}
    \caption{Rest-mass fraction of the ejected mass $M_{\rm ejecta}$ following the GW 
    peak amplitude computed in the domain $r>30M$.}
    \label{fig:ejecta-mass}
\end{figure}

Figure~\ref{fig:ejecta-mass} shows the dynamically ejecta rest-mass fraction of the unbound material 
computed via Eq.~\eqref{eq:ejecta-mass}. We note that depending on the magnetic field topology the ejecta ranges from $\sim0.5\%$ and $\sim3.0\%$ of the total mass of the system, corresponding to $M_{\rm eje}\sim 10^{-2}-10^{-1} \,\Msol$ (see Table~\ref{tab:results}), with average velocity $\langle v_{\rm eje}\rangle\sim0.2-0.3\,c$. Ejecta masses $\gtrsim10^{-3}\Msol$ are expected to lead to detectable, transient kilonova signatures~\cite{Metzger:2019zeh} originating from the decay of 
radioactive nuclei synthesized via the r-process within the neutron-rich material ejected during BNS mergers. 

We note that, regardless of the magnetic field configurations, the dynamical ejection of matter 
for the SLy binaries begins $\sim 1\,{\rm ms}$ before that from the WFF1 ones. This is expected, 
since the SLy binaries are less compact than WFF1 ones, and therefore, they experience stronger tidal 
deformation before merger. This deformation stretches the outer layers even before contact, 
priming them for earlier dynamical ejection. However, we observe that near the termination of our 
simulations, the soft-WFF1-EoS binaries exhibit a larger overall ejection of material. This may be 
attributed to the higher compactness of WFF1 stars, which leads to a more violent merger interface, driving enhanced mass ejection 
at later times.  

On the other hand, we observe that the dynamical ejection of matter in binaries with {\bf{T}$_{0.95}$}, begin slightly earlier
than the {\bf{P}} configurations.  Toroidal fields, being confined to the stellar interior, enhance local magnetic pressure and promote stronger compression and shock heating during merger~\cite{Kiuchi:2014hja}. Additionally, they impose less magnetic tension opposing radial outflows compared to poloidal fields, allowing earlier  development of dynamical ejecta \cite{Radice:2024gic}.

To gain insight into the possible EM signatures from the dynamical ejection of matter, 
we employ the analytical model presented in~\cite{Perego:2021dpw}. This model computes the 
peak rise times, bolometric luminosities and the effective temperatures of the kilonova 
assuming that the ejecta is spherically distributed and expanding homologously with an 
average speed $v_{\rm eje}$ and characterized by a gray opacity $\kappa_\gamma$. 

The peak time of the transient emission can be estimated as
\begin{align} \label{eq:kilonova_t}
    \tau_{\rm peak} &\sim \sqrt{\frac{M_{\rm eje}\kappa_\gamma}{4\pi\langle v_{\rm eje}\rangle c}} \nonumber \\
    &\approx 4.6 \, {\rm days}\, \left(\frac{M_{\rm eje}}{10^{-2}\Msol}\right)^{1/2}\left(\frac{\langle v_{\rm eje}\rangle}{0.1c}\right)^{-1/2} \,.
\end{align}
We assume a high opacity $\kappa_\gamma=10\,{\rm cm}^2{\rm g}^{-1}$, which corresponds to ejecta containing 
an electron fraction $Y_e\lesssim 0.25$, a typical value expected from the early post-merger phase in 
BNSs~\cite{Fujibayashi:2022ftg}.
The peak luminosity can be computed as
\begin{equation} \label{eq:kilonova_L}
    L_{\rm knova} \sim 2.4\times10^{40}\left(\frac{M_{\rm eje}}{0.01\Msol}\right)^{0.35}
    \left(\frac{\langle v_{\rm eje}\rangle}{0.1c}\right)^{0.65} {\rm erg\,s}^{-1} \,.
\end{equation}
Finally,  assuming black body emission, and using the Stefan-Boltzmann law, the effective 
temperature at the peak can be be estimated as
\begin{equation} \label{eq:kilonova_T}
    T_{\rm peak} \sim 2.15\times10^3\left(\frac{M_{\rm eje}}{0.01\Msol}\right)^{-0.16}
    \left(\frac{\langle v_{\rm eje}\rangle}{0.1c}\right)^{-0.09} {\rm K} \,.
\end{equation}
While simplifying, these assumptions provide a useful first-order model that generally 
fits observed data. For our simulations, we find that the bolometric luminosity is 
$L_{\rm knova}\sim10^{40.6\pm 0.1}\,{\rm erg\,s}^{-1}$ with rise times of $\tau_{\rm peak} 
\sim 2.9-9.4$ days and an effective temperature of $T_{\rm peak} \sim 10^{3.3} \,{\rm K}$,
which is consistent with those obtained from other numerical simulations ~\cite{Ruiz:2021qmm, Bamber:2024kfb} (see \cite{Metzger:2019zeh} for a review).
This temperature can be converted into a peak wavelength
$\lambda_{\rm peak}\sim 1.35\times 10^{3}\,{\rm nm}\,(T_{\rm peak}/10^{3.3}\,\rm K)^{-1}$, and the emission can be detected with
current or planned telescopes, such as ALMA or the Vera C. Rubin observatory~\cite{2018PASP..130a5002M}.
As expected, we find that EM signatures (Poynting and kilonova luminosities) depend more heavily on the magnetic field topology than on the EoS.

\section{Conclusions}
\label{sec:conclusions}

Multimessenger detections of electromagnetic signals and gravitational waves, such as GW170817 and its EM counterparts, have proven crucial in improving our understanding of the sources of these signals. Future detections will allow us to further constrain the EoS of neutron stars and to understand their magnetic fields. With the improved sensitivity of future detectors in sight, we expect an increased quality and quantity of data to study these systems. Therefore, it is of the utmost importance to model these events beforehand, in order to accurately recover the properties of their sources.

The interior magnetic field of neutron stars is, in particular, one of the big questions surrounding these objects. Motivated by observations of pulsars' magnetic fields, GRMHD simulations of BNS mergers typically consider a dipole pulsar-like structure for the magnetic field, usually confined to the interior of the stars. However, their interior field remains unknown. In this study, we consider four distinct magnetic field topologies: a dipole pulsar-like field ({\bf P}), a linear superposition of poloidal and toroidal components inside the star ({\bf SP}), and a confined tilted toroidal component transitioning to a pulsar-like topology at radii $0.95\,R_{\rm NS}$ ({\bf T$_{0.95}$}) and $0.5\,R_{\rm NS}$ ({\bf T$_{0.5}$}). We analyze their evolution, EM and GW signals, as well as, if applicable, their remnant properties. We consider two different equations of state, SLy and WFF1, and fix the maximum value of $\beta^{-1}\equiv P_{\rm mag}/P_{\rm gas}=0.003125$ inside the stars. Additionally, we consider a lower-mass system with the SLy EoS, a pulsar-like field and $\beta^{-1}=0.0023$, in order to confirm the same behavior we observe in the WFF1 remnants, as all other SLy simulations lead to a black hole.

We perform an extensive multimessenger analysis of the emission properties of the systems, both gravitational waves and electromagnetic signatures. We study the main active modes that contribute to gravitational emission and their corresponding density eigenfunctions. We further analyze the properties of the remnants, namely their rotational profile, temperature distribution, and convective stability. 

Our main results can be summarized as follows:
\begin{itemize}
    \item Magnetic field topology significantly impacts GW emission, altering both the collapse time and the $f_2$ peak frequency in longer-lived remnants;
    \item Frequency shifts in the early $f_{2i}$ mode due to different magnetic fields will be detectable with third-generation detectors up to $\sim50\,{\rm Mpc}$;
    \item Longer-lived remnants exhibit additional frequency modes beyond the fundamental $l=m=2$, some identified through an extra radial node in their eigenfunctions. In the pulsar-like case, this mode is consistent with a non-linear coupling between the $m=0$ and $m=2$ modes;
    \item All configurations produce Poynting luminosities consistent with sGRBs, and are compatible with kilonova emission;
    \item A purely poloidal configuration is the most efficient in launching a relativistic jet;
    \item Hot rings form at the centrifugal barrier and correlate with regions of convective stability.
\end{itemize}

Finally, we stress that future work should incorporate neutrino transport, more advanced magnetic topologies, and finite-temperature EoSs to refine the multimessenger predictions and clarify the role of NSs as potential sGRB progenitors.

\begin{acknowledgments}
We thank Pablo Cerdá-Durán and Dimitra Tseneklidou for useful discussions. 
We also thank members of our Illinois Relativity Undergraduate Research Team,
in particular, Eric Yu for his help with some of the 3D visualizations. 
This work was supported in part by the Generalitat Valenciana (grants CIDEGENT/2021/046 
and Prometeo CIPROM/2022/49), and by the Spanish Agencia Estatal de Investigación 
(grants PID2024-159689NB-C21 funded by the Ministerio de Ciencia, Innovación y Universidades, PID2021-125485NB-C21 funded by MCIN/AEI/10.13039/501100011033 
and ERDF A way of making Europe). By the National Science Foundation (NSF)
Grants PHY-2308242, OAC-2310548 and PHY-2006066 to the University of Illinois at Urbana-Champaign. DG acknowledges support from the Spanish Agencia Estatal de  Investigaci\'on through the grant "Ayuda para la Formación del Personal investigador" (FPI) fellowship No.~PRE2019-087617. MMT acknowledges support from the Science and Technology Facilities Council (STFC), via grant 
No.~ST/Y000811/1, and from the Ministerio de Ciencia, Innovación y Universidades del 
Gobierno de España through the “Ayuda para la Formación de Profesorado Universitario" (FPU) fellowship No.~FPU19/01750. A.T. acknowledges support from the National Center for Supercomputing Applica-
tions (NCSA) at the University of Illinois at Urbana-Champaign through the NCSA Fellows program. 
This work used Frontera at the Texas Advanced Computing Center (TACC) through allocation AST20025, 
MareNostrum 5 at the Barcelona Supercomputing Center allocations AECT-2025-2-0010 and AECT-2025-2-0031.
The authors acknowledge the computational resources provided by the LIGO Laboratory and supported
by National Science Foundation Grants PHY-0757058 and PHY-0823459, and the resources from the Gravitational 
Wave Open Science Center, a service of the LIGO Laboratory, the LIGO Scientific Collaboration and the Virgo Collaboration.
\end{acknowledgments}

\appendix
\section{Magnetic Power Spectrum} \label{appendix:magnetic-power-spectrum}
\begin{figure}
    \centering
    \includegraphics[width=0.95\linewidth]{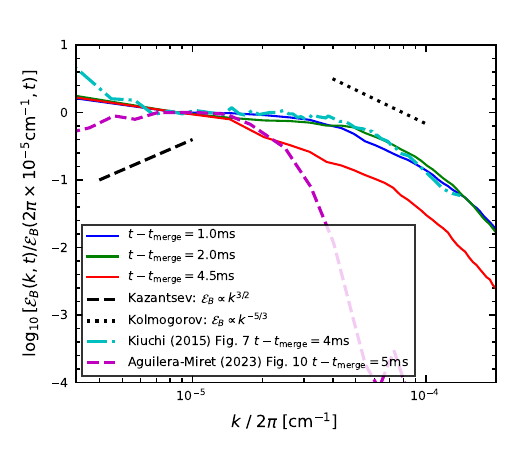}
    \caption{Magnetic power spectrum computed via \cite{Aguilera-Miret:2020dhz} Eq. B1, as used in \cite{Aguilera-Miret:2023qih}, at different times post-merger for the WFF1\_P simulation.}
    \label{fig:power-spectrum-WFF1_P}
\end{figure}

\begin{figure}
    \centering
    \includegraphics[width=0.95\linewidth]{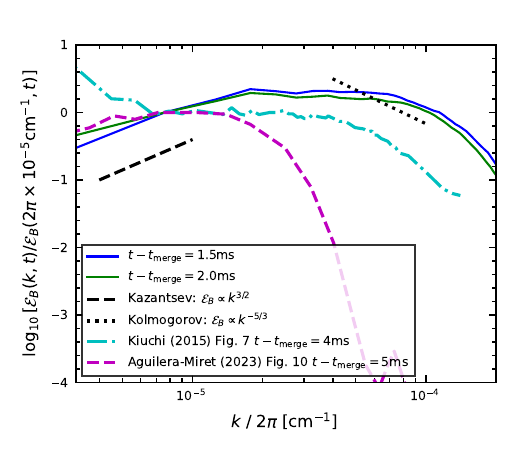}
    \caption{Magnetic power spectrum computed via \cite{Aguilera-Miret:2020dhz} Eq. B1, as used in \cite{Aguilera-Miret:2023qih}, at different times post-merger for the SLy\_SP simulation.}
    \label{fig:power-spectrum-SLy_SP}
\end{figure}

To quantify the development of magnetohydrodynamic turbulence in the post-merger remnant, we compute the magnetic power spectrum following Eq. (B1) of \cite{Aguilera-Miret:2020dhz}. The magnetic power spectrum $\varepsilon_B(k,t)$ characterizes the distribution of magnetic energy across spatial scales and provides insight into the dominant physical processes driving magnetic field amplification and energy cascade. Specifically, it allows us to assess whether the post-merger magnetic field growth is governed by a small-scale dynamo (e.g., Kazantsev spectrum, $\varepsilon_B \propto k^{3/2}$) or fully developed turbulence (e.g., Kolmogorov spectrum, $\varepsilon_B \propto k^{-5/3}$).
We present the results for two specific simulations, involving the two different EoSs and two distinct magnetic field configurations. In Fig.~\ref{fig:power-spectrum-WFF1_P}, the magnetic power spectrum of WFF1\_P is consistent with previous high-resolution studies \cite{Kiuchi:2015sga}, especially for earlier times post-merger. At smaller scales ($k/2\pi\gtrsim 7\times10^{-4}$), the spectra resemble a Kolmogorov-like slope, and at later times they do so for $k/2\pi\gtrsim 10^{-5}$, indicating a transition toward turbulent energy redistribution across scales. In Fig.~\ref{fig:power-spectrum-SLy_SP}, the magnetic power spectrum of SLy\_SP shows growth, for $k/2\pi\lesssim 9\times10^{-4}$, with slopes compatible with the Kazantsev regime, suggesting amplification via small-scale dynamo action.
The precise shape and evolution of the power spectrum remain sensitive to numerical resolution, filtering methods used to compute the Fourier transform, and the physical parameters of the system. Further studies are needed to robustly determine the spectrum's behavior and its dependence on such parameters. Nevertheless, our results are in broad agreement with higher-resolution and large-eddy studies \cite{Kiuchi:2015sga, Aguilera-Miret:2023qih}.

\section{Convective Stability} \label{appendix:convective-stability}
The Brunt-V\"ais\"ala frequency $~\mathcal{N}$ quantifies the oscillation frequency of a displaced fluid element around its equilibrium position \cite{Cox80, Tassoul_2000}. It is closely associated with the Ledoux discriminant $\mathcal{B}_i$, whose sign indicates the convective stability of a given region \cite{Cox80, Boston:2023fey}. Therefore, we can compute $\mathcal{N}^2$ to understand what regions are convectively stable/unstable. In particular, a region with a positive $\mathcal{N}^2$ is stable against convection and propagation of g-modes. Furthermore, it provides an estimate of the frequency of the g-modes. We have that $\mathcal{N}^2\propto \mathcal{B}_i\mathcal{G}^i$ \cite{Cerda-Duran:2007fvm, Torres-Forne:2018nzj, Gao:2025nfj}, where $\mathcal{B}_i$ is the relativistic version of the Ledoux discriminant, and $\mathcal{G}_i$ is the gravitational acceleration. The proportionality factor is coordinate dependent. In the conformally flat condition approximation, this factor is $\alpha^2/\psi^4$ \cite{Torres-Forne:2018nzj}, where $\alpha$ is the lapse function and $\psi$ the conformal factor. For neutron stars, both $\alpha$ and $\psi$ are of the order units, as can be seen from post-Newtonian analysis \cite{Blanchet:1989fg}. On the other hand, considering the metric for an axisymmetric rotating star, as was recently done in \cite{Gao:2025nfj}, this factor depends on $\gamma_{rr}$, and more generally on $g_{rr}$. In either case, the proportionality factor is expected to be $\sim \mathcal{O}(1)$, which differs the most from unity at the core where the gravitational field is the strongest. We have
\begin{equation}\label{eq:Bi}
    \mathcal{B}_i \equiv \frac{\partial_i e}{\rho_0 h} - \frac{1}{\Gamma_1} \frac{\partial_i P}{P} \, ,
\end{equation}
with $e$ the energy density, $h$ the specific enthalpy, $\Gamma_1$ the adiabatic index around a pseudo-barotropic equilibrium, $\rho_0$ the rest-mass density and $P$ the pressure. Additionally, the gravitational acceleration is given by $\mathcal{G}_i=-\partial_i\ln\alpha$, with $\alpha$ the lapse function. In hydrostatic equilibrium we can also write \cite{Torres-Forne:2018nzj}
\begin{equation}\label{eq:Gi}
    \mathcal{G}_i \equiv -\partial_i \ln \alpha = \frac{1}{\rho_0 h}\partial_i P \, .
\end{equation}
We compute $\mathcal{N}^2$ in the equatorial plane using cylindrical coordinates. Simulations of core-collapse supernovae have shown that the two different definitions of $\mathcal{G}_i$ in Eq.~\eqref{eq:Gi} differ at most by a factor of $2$ \cite{Torres-Forne:2018nzj} (see the lower right panel of their Figure 11).
In Fig.~\ref{fig:tempvsbrunt} and~\ref{fig:tempvsbruntSLy}, we compute $\mathcal{N}$ via
\begin{align} \label{eq:brunt}
    \mathcal{N} &= \mathcal{N}^2/|\mathcal{N}|\, , \quad {\rm with} \\ \label{eq:bruntsq}
    \mathcal{N}^2 &\approx \left(\frac{\partial_r e}{\rho_0 h} - \frac{1}{\Gamma_1} \frac{\partial_r P}{P}\right) \frac{\partial_r P}{\rho_0 h} \, .
\end{align}
We can use it as a proxy for convective stability. However, its values should be interpreted with caution, as they represent only rough estimates of the true value due to the approximations employed in this work. Furthermore, a complete study of convective stability in magnetized remnants should consider dynamo mechanisms, such as the $\alpha$-$\Omega$-dynamo and Tayler-Spruit dynamo. We remark that a detailed study of convective stability is beyond the scope of this work.

\bibliographystyle{apsrev4-1}
\bibliography{ref}

\end{document}